\documentclass[aps,prc,preprint,eqsecnum,showpacs,floatfix]{revtex4-1}

\usepackage{graphicx,epstopdf}
\usepackage{hyperref}
\usepackage{color}
\usepackage{subfigure}
\usepackage{amsmath,mathtools}
\usepackage{amssymb}
\usepackage{epsfig}
\usepackage{amsfonts}
\usepackage{bbm}
\usepackage{bm}
\usepackage{float}
\usepackage{xcolor}

\def\bra#1{\bigl\langle{ #1} \bigr|}
\def\ket#1{\bigl|{ #1} \bigr\rangle}

\def\ovlp#1#2{\bigl\langle{ #1}\big|{#2} \bigr\rangle}

\def\pd{{\phantom{\dagger}}}
\def\rvec {{\bf r}}
\def\pvec {{\bf p}}
\def\hvec {{\bf h}}
\def\qvec {{\bf q}}
\def\kvec {{\bf k}}
\def\creat#1{a_{#1}^{\dagger}}                    
\def\annil#1{a_{#1}^{\pd}}                      

\def\1{\mathbbm{1}}
\def\EF{e_{\rm F}}
\def\KF{k_{\rm F}}
\def\SF{S_{\rm F}}
\def\a0{a_0}
\def\I{{\rm i}}

\def\Im{{\cal I}}

\def\ie{{\em i.e.\/}\ }

\makeatletter
\def\mathcenterto#1#2{\mathclap{\phantom{#1}\mathclap{#2}}\phantom{#1}}
\let\old@widetilde\widetilde
\def\widetildeto#1#2{\mathcenterto{#2}{\old@widetilde{\mathcenterto{#1}{#2\,}}}}
\let\old@widehat\widehat
\def\widehatto#1#2{\mathcenterto{#2}{\old@widehat{\mathcenterto{#1}{#2\,}}}}
\makeatother

\begin{document}

\title{Variational and parquet-diagram calculations for neutron
  matter.\\ III.  S-wave pairing}

\author{E.~Krotscheck$^{\dagger\ddagger}$ and J. Wang$^\dagger$}

\affiliation{$^\dagger$Department of Physics, University at Buffalo, SUNY
Buffalo NY 14260}
\affiliation{$^\ddagger$Institut f\"ur Theoretische Physik, Johannes
Kepler Universit\"at, A 4040 Linz, Austria}

\begin{abstract}
  We apply parquet-diagram summation methods for the calculation of
  the superfluid gap in $S$-wave pairing in neutron matter for
  realistic nucleon-nucleon interactions such as the Argonne $v_6$ and
  the Reid $v_6$ potentials. It is shown that diagrammatic
  contributions that are outside the parquet class play an important
  role. These are, in variational theories, identified as so-called
  ``commutator contributions''. Moreover, using a particle-hole
  propagator appropriate for a superfluid system results in the
  suppression of the spin-channel contribution to the induced
  interaction. Applying these corrections to the pairing interaction,
  our results agree quite well with Quantum Monte Carlo data.
\end{abstract}
\maketitle

\section{Introduction} 
\label{sec:intro}

The nature and role of fermionic pairing and superfluidity in nuclei
and nuclear matter has been a subject of great interest for many years
\cite{BCS50book}.  Beginning with work by Bohr, Mottelson, and Pines
\cite{BMP} there was persistent interest among nuclear theorists in
what could be learned from the quantum many-body problem of infinite
nuclear matter composed of nucleons interacting through the best
nucleon-nucleon (NN) interaction available.

BCS theory as originally formulated \cite{BCS} is intrinsically a mean
field theory. Cooper, Mills, and Sessler \cite{CMS} were the first to
realize that the BCS equation {\em per se\/} could also be solved for
hard-core interactions, but that still leaves the question open to
what extent such a theory could capture the physics of a strongly
interacting system. This issue was addressed by the introduction of
Jastrow-Feenberg correlation factors
\cite{YangNC,YangClarkBCS,YangThesis}. Major advances were made with
the replacement of cluster expansions by Fermi hypernetted-chain
(FHNC) diagram-resummation techniques \cite{Johnreview,KroTrieste},
facilitating the unconstrained optimization of Jastrow-Feenberg
correlations (FHNC-EL method). The fact that optimized
hypernetted-chain summations included the summations of high-order
contributions to the perturbation series was first observed by Sim,
Buchler, and Woo \cite{Woo70}, it was put on a rigorous foundation in
the work by Jackson, Lande, and Smith \cite{parquet1,parquet2} who
showed, for bosons, that the optimized hypernetted chain theory for
Jastrow-Feenberg correlations is equivalent to the self-consistent
summation of all ring- and ladder diagrams, the so-called ``parquet''
diagrams.

When implemented in a BCS extension, these advances have made possible
the development of a rigorous correlated BCS (CBCS) theory
(\citenum{CBFPairing}, see also Ref.~\citenum{HNCBCS}) that respects the
U(1) symmetry-breaking aspect of the superfluid state -- \ie
the non-conservation of particle number.  A recent in-depth study of
correlations in the low-density Fermi gas \cite{cbcs}, with emphasis on
the presence of Cooper pairing and dimerization, documents the power
of the Euler-Lagrange FHNC approach adopted in the present work.
The major drawback of these calculations was that they employed
simple state-independent correlation functions. This makes the
method suitable for simple interactions, but improvements must be sought
for realistic nuclear Hamiltonians.

In recent work, \cite{v3eos,v3twist} we have utilized the equivalence
between parquet-diagram summations and optimized variational methods
to develop methods that address exactly this problem. We will review
these in the next section.

\section{Variational and parquet-diagram theory}
\label{sec:GMBT}
\subsection{The normal ground state}
\label{ssec:FHNC}

Let us briefly describe the Jastrow-Feenberg variational
and parquet-diagram summation method and its implementation to
superfluid systems.

We assume a
non-relativistic many-body Hamiltonian
\begin{equation}
H = -\sum_{i}\frac{\hbar^2}{2m}\nabla_i^2 + \sum_{i<j}
v(i,j)\,.
\label{eq:Hamiltonian}
\end{equation}
Popular models of the nucleon-nucleon force
\cite{Reid68,Bethe74,Day81,AV18,Wiri84} represent the interaction as a
sum of local functions times correlation operators, \ie
\begin{equation}
\hat v (i,j) = \sum_{\alpha=1}^n v_\alpha(r_{ij})\,
        \hat O_\alpha(i,j),
\label{eq:vop}
\end{equation}
where $r_{ij}=|\rvec_i-\rvec_j|$ is the distance between particles $i$
and $j$, and the $O_\alpha(i,j)$ are operators acting on the spin,
isospin, and possibly the relative angular momentum variables of the
individual particles.  According to the number of operators $n$, the
potential model is referred to as a $ v_n$ model potential. Reasonably
realistic models for nuclear matter keep at least the six base
operators, these are
\begin{eqnarray}
\hat O_1(i,j;\,\hat\rvec_{ij})
        &\equiv& \hat O_c = \1\,,
\nonumber\\
\hat O_3(i,j;\,\hat\rvec_{ij})
        &\equiv& {\bm\sigma}_i \cdot {\bm\sigma}_j\,,
\nonumber\\
\hat O_5(i,j;\,\hat\rvec_{ij})
&\equiv& S_{ij}(\hat\rvec_{ij})
      \equiv 3({\bm\sigma}_i\cdot \hat\rvec_{ij})
      ({\bm\sigma}_j\cdot \hat\rvec_{ij})-{\bm\sigma}_i \cdot {\bm\sigma}_j\,,
      \nonumber\\
      \hat O_{2n}(i,j;\,\hat\rvec_{ij}) &=& \hat O_{2n-1}(i,j;\hat\rvec_{ij})
      {\bm\tau}_1\cdot{\bm\tau}_2\,,
\label{eq:operator_v6}
\end{eqnarray}
where $\hat\rvec_{ij} = \rvec_{ij}/r_{ij}$. We will omit the
arguments when unambiguous.

There are basically two methods of comparable diagrammatic richness
for manifestly microscopic calculations of properties of such strongly
interacting systems. These are the Jastrow-Feenberg variational method
\cite{FeenbergBook} and the parquet-diagram summations
\cite{parquet1,parquet2}. For Bose systems, and for purely central
interactions, these two methods have been shown to lead to exactly the
same equations. For a strongly interacting and translationally invariant {\em
normal\/} system, the Jastrow-Feenberg  method starts with an {\em
  ansatz\/} for the wave function, \cite{FeenbergBook}
\begin{eqnarray}
\Psi_0({\bf r}_1,\ldots,{\bf r}_N) &=&
	F({\bf r}_1,\ldots,{\bf r}_N)
	\Phi_0(1,\ldots,N)\label{eq:wavefunction},\\
F({\bf r}_1,\ldots,{\bf r}_N) &=& \prod^N_{\genfrac{}{}{0pt}{2}{i,j=1}{i<j}}f(r_{ij})\label{eq:Jastrow}
\end{eqnarray}
where $\Phi_0({\bf r}_1,\ldots,{\bf r}_N)$ denotes a model state,
which for normal Fermi systems is a Slater-de\-ter\-mi\-nant, and $F$
is the correlation operator which can, of course, also contain
three-body correlations. For Bose systems, $\Phi_0(1,\ldots,N)=1
$. The correlation functions $f(r_{ij})$ are obtained by minimizing
the energy, {\em i.e.\/} by solving the Euler-Lagrange (EL) equations
\begin{eqnarray}
 E_0 && = \frac{\left\langle\Psi_0\right|H\left|\Psi_0\right\rangle}
  {\left\langle\Psi_0\mid\Psi_0\right\rangle}
\equiv H_{{\bf o}}\label{eq:energy}\,,\\
&&        \frac{\delta E_0}
{\delta f}(r_{12}) = 0\,.
\label{eq:euler}
\end{eqnarray}

Evaluation of the energy (\ref{eq:energy}) for the variational wave
function (\ref{eq:wavefunction}, \ref{eq:Jastrow}) and analysis of the
variational problem are carried out by cluster expansion and
resummation methods. The procedure has been described at length in
review articles \cite{Johnreview,polish} and pedagogical material
\cite{KroTrieste}.

No derivation comparable in rigor to that of
Refs. \onlinecite{parquet1,parquet2} exists for fermions. We have
analyzed in Ref. \onlinecite{fullbcs} the relationship between
specific classes of diagrams generated by the cluster expansion and
optimization procedure of Jastrow-Feenberg theory, and classes of
parquet diagrams, specifically rings, ladders, and self-energy
corrections.  Besides the localization procedures used to establish
the agreement between the boson versions of Jastrow-Feenberg and
parquet diagrams, a ``collective'' approximation must be made for the
particle-hole propagator. Moreover, since the Fermi sea breaks
Galilean invariance, specific Fermi sea averages must be made to make
all two-body vertices functions of the momentum transfer only. These
procedures have been discussed and examined in detail in
Ref. \onlinecite{fullbcs}.

The situation is much more complicated for realistic nuclear
Hamiltoninans of the form \eqref{eq:vop}. A plausible generalization
of the Jastrow-Feenberg function \eqref{eq:Jastrow} would be
\cite{Pand76,FantoniSpins,IndianSpins} the so-called ``symmetrized
operator product form
\begin{equation}
        \Psi_0^{{\rm SOP}}
        = {\cal S} \Bigl[ \prod^N_{\genfrac{}{}{0pt}{2}{i,j=1}{i<j}} \hat f (i,j)\Bigr] \Phi_0\,,
\label{eq:f_prodwave}
\end{equation}
where
\begin{equation}
  \hat f(i,j) = \sum_{\alpha=1}^n f_\alpha(r_{ij})\,
  \hat O_\alpha(i,j)\,,
  \label{eq:fop}
\end{equation}
and ${\cal S}$ stands for symmetrization. The symmetrization is
necessary because the operators $\hat O_\alpha(i,j)$ and $\hat
O_\beta(i,k)$ do not necessarily commute. The need to symmetrize the
operator product causes, however, severe complications and so far no
summation that comes anywhere close to the diagrammatic richness of
the (F)HNC summations for state-independent correlations has been
found. As a consequence, no unconstrained optimization method
analogous to Eq. \eqref{eq:euler} could be developed.  Instead, the
correlation functions $f_\alpha(r)$ have been either assumed to be of
some simple parameterized form, or calculated by a low-order effective
Schr\"odinger equation (``low order constrained variation'',
LOCV). Operator contributions were calculated in a chain approximation
``single-operator chains (SOC)'' which can be understood \cite{rings}
as a simplified version of the random phase approximation (RPA). We
have shown in previous work \cite{SpinTwist} that this leads to
sensible results only if the so-called commutator terms generated by
the symmetrization of the correlation operator \eqref{eq:fop} are
omitted.

In view of these complications, Smith and Jackson \cite{SmithSpin}
developed the parquet-diagram summations for a fictitious system of
bosons interacting via a $v_6$ model Hamiltonian. It turned out that
the equations derived were the same as the Bose version of the
hypernetted chain equations derived from a variational wave function
(\ref{eq:f_prodwave},\ref{eq:fop}) when all commutators are omitted,
and supplemented by the optimization condition \eqref{eq:euler}. This
leads to the conclusion that the commutator diagrams correspond to
diagrams in perturbation theory that are beyond the parquet class.

\begin{figure}
  \centerline{\includegraphics[width=0.5\columnwidth]{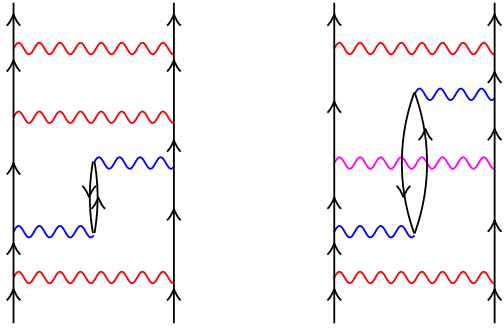}}
  \caption{(color online) The figure shows the essential processes are
    included in the ``twisted chain'' interaction correction. The red
    wavy lines are either spin-singlet or spin-triplet interactions,
    the magenta line may be either of the two, and the chain of blue
    lines represents a contribution to the induced interaction
    $\widehat{W}_{\rm I}$ (from
    Ref. \onlinecite{v3twist}).\label{fig:vsummary}}
  \end{figure}

The physical mechanism described by commutator diagrams is exemplified
in the two simple processes shown diagrammatically in
Fig. \ref{fig:vsummary}. In the left diagram, a pair of particles that
enter the process in a specific (singlet or triplet) state will always
remain in that state. The red wavy lines therefore describe
interactions in the {\em same\/} channel. This is not changed by the
exchange of a (spin-)density fluctuation depicted by the chain of two
blue lines. In the right diagram, a spin is absorbed, transported
through a spin-fluctuation, described again by the chain of two blue
wavy lines, and re-absorbed at a later time. In that situation, the
magenta wavy line may be a triplet interaction whereas the red lines
are singlet interactions or vice versa. Evidently, this makes little
difference if the interactions are the same in spin-singlet and
spin-triplet states. On the other hand, there is no reason that the
two processes are similar if the interactions are very different,
which is the case for modern nucleon-nucleon interactions
\cite{Reid68,AV18}.

Taking this into account and the evidence that simplistic choices of
the pair correlation functions $f_\alpha(r_{ij})$ lead to sensible
results only when commutator diagrams are omitted, we have in recent
work \cite{v3twist} added the leading corrections that capture the
essential physics of the commutator diagrams. To make the method
practical, we have used approximations suggested by the
Jastrow-Feenberg theory and the insight about diagram topology from
parquet diagram summations. The results showed that the
``beyond-parquet'' diagrams are, especially in low density neutron
matter and in the singlet interaction channel, more important than any
other many-body corrections.

\subsection{Strongly interacting superfluids}
\label{ssec:cbcs}

Let us now turn to the generalization of the correlated wave functions
method to superfluid systems. Having reviewed the FHNC-EL theory and
its relation to parquet diagrams above, we can restrict ourselves to
the discussion of what changes for a superfluid system.  Older work
has either assumed that the superfluid state deviates little from the
normal state \cite{ectpaper,HNCBCS,CBFPairing,shores,CCKS86,cbcs}
and/or adopted low-order cluster expansions
\cite{YangClarkBCS,Fabrocinipairing,Pavlou2017,Benhar}. In recent work
\cite{fullbcs}, we have developed the Jastrow-Feenberg variational
approach for a superfluid system to a level comparable to that of the
normal system. This has made the identification with parquet-diagrams
possible. A number of important results will be discussed below.

The basic idea of a correlated BCS state is to use for the model state
in Eq. \eqref{eq:wavefunction} an uncorrelated
BCS state
\begin{equation}
\ket{\rm BCS} =
{\prod_{\kvec}}
\left[ u_{\kvec} +  v_{\kvec} a_{ {\bf  k} \uparrow }^\dagger
 a_{-{\bf  k} \downarrow}^\dagger  \right] \ket{\,}\,
\label{eq:BCS}
\end{equation}
where $\ket{\,}$ is the vacuum state and the $u_\kvec$, $v_\kvec$ are
the Bogoliubov amplitudes satisfying $u_\kvec^2 + v_\kvec^2 = 1$. A
{\em correlated\/} state is then constructed by applying a correlation
operator $F$ to that state. Since the state \eqref{eq:BCS} does not
have a fixed particle number, we must write the correlated state in
the form
\begin{equation}
\ket{\rm CBCS} =  \sum_{{\bf m},N} \ket {\Psi_{\bf m}^{(N)}}
\ovlp{{\bf m}^{(N)}}{\rm BCS}
\label{eq:CBCS}
\end{equation}
where the $\{\ket{{\bf m}^{(N)}}\}$ form a complete set of $N$-body Slater
determinants, and the $\ket {\Psi_{\bf m}^{(N)}}$ are correlated and normalized
$N$-body states forming a non-orthogonal basis of the Hilbert space,
see Eq. \eqref{eq:States}.

In what follows, we will refer to expectation values with respect to
the {\em uncorrelated\/} state (\ref{eq:BCS}) as
$\left\langle\ldots\right\rangle_0$ and those with respect to the {\em
  correlated state\/} (\ref{eq:CBCS}) as
$\left\langle\ldots\right\rangle_c$.  Physically interesting
quantities like the (zero temperature) Landau potential of the
superfluid system
\begin{equation}
  \left\langle H'\right\rangle_c = \frac{\bra{\mathrm{CBCS}} \hat H'
    \ket{\mathrm{CBCS}}}
  {\bigl\langle\mathrm{CBCS}\big|\mathrm{CBCS}\bigr\rangle}\,,\qquad
  \hat H' \equiv \hat H - \mu\hat N\,.
  \label{eq:EBCS}
\end{equation}
are then calculated by cluster expansion and resummation techniques.
Above, $\mu$ is the chemical potential.

There are basically two ways to deal with the correlated wave function
\eqref{eq:CBCS}.

\subsubsection{Weakly coupled systems}
\label{ssec:weak}

We rely in this section heavily on definitions and methods of
correlated basis functions (CBF) theory that have been discussed
elsewhere \cite{Johnreview,polish,KroTrieste}. To settle the notation,
we give the definitions of the essential quantities in Appendix
\ref{app:CBF}.

If the superfluid gap is small compared to the Fermi energy, it is
legitimate to simplify the problem by expanding $\left\langle
H'\right\rangle_c$, Eq. \eqref{eq:EBCS} in the {\em deviation\/} of
the Bogoliubov amplitudes $u_{\kvec}$, $v_{\kvec}$ from their normal
state values $u^{(0)}_{\kvec}= \bar n(k)$, $v^{(0)}_{\kvec}=n(k)$,
where $n(k) = \theta(\KF-k)$ is the Fermi distribution and $\bar n(k)
= 1-n(k)$. This approach adopts a rather different concept than the
original BCS theory: A wave function of the form (\ref{eq:BCS}) begins
by creating Cooper pairs out of the vacuum. Instead, the approach
(\ref{eq:CBCS}) begins with the {\em normal, correlated\/} ground
state and generates one Cooper pair at a time out of the normal system
as suggested recently by Leggett \cite{LeggettQFS2018}. Adopting such
an expansion in the number of Cooper pairs, the correlation functions
$f(r_{ij})$ and possibly higher order correlations can be optimized
for the normal system.

Carrying out this expansion in the number of Cooper pairs, we have
arrived in Ref. \citenum{CBFPairing} at the energy expression of the
superfluid state
\begin{eqnarray}
\langle \hat H' \rangle_c &=& E_0 - \mu N + 2 \sum_{\kvec
  ,\,|\, \kvec \,|\,>\KF} v_{\kvec}^2 (e_{\kvec} - \mu ) - 2
\sum_{\kvec , \,|\, \kvec \,|\,<\KF} u_{\kvec}^2 (e_{\kvec} - \mu )
\nonumber \\ &\quad& + \sum_{\kvec,\kvec'}u_\kvec v_\kvec u_{\kvec'}
v_{\kvec'} {\cal P}_{\kvec\kvec'}\,.
\label{eq:Ebcs}
\end{eqnarray}
Above, $E_0\equiv H_{\bf o}^{(N)}$ is the energy expectation value
\eqref{eq:energy} of the normal $N$-particle system.  The $e_{\kvec}$
are the single particle energies derived in CBF theory \cite{CBF2},
see Appendix \ref{app:CBF}. The paring interaction has the form
\begin{eqnarray}
{\cal P}_{\kvec\kvec'} &=& {\cal W}_{\kvec\kvec'}+(|e_{\kvec}- \mu | 
+ |e_{\kvec'}- \mu |)
{\cal N}_{\kvec\kvec'}\label{eq:Pdef}\,,\\
{\cal W}_{\kvec\kvec'} &=& \bra{\kvec \uparrow ,-\kvec\downarrow}
{\cal W}(1,2)\ket{\kvec'\uparrow ,-\kvec'\downarrow}_a\,,\label{eq:Wnldef}\\
{\cal N}_{\kvec\kvec'}&=&
\bra{\kvec \uparrow ,-\kvec\downarrow}
{\cal N}(1,2)\ket{\kvec'\uparrow , - \kvec'\downarrow}_a\,.
\label{eq:Ndef}\end{eqnarray}
The effective interaction ${\cal W}(1,2)$ and the correlation
corrections ${\cal N}(1,2)$ are given by the compound-diagrammatic
ingredients of the FHNC-EL method for off-diagonal quantities in CBF
theory \cite{CBF2}.

The Bogoliubov amplitudes $u_\kvec $, $v_\kvec $ are obtained in the
standard way by variation of the energy expectation (\ref{eq:Ebcs}).
This leads to the familiar gap equation
\begin{equation}
\Delta_\kvec = -\frac{1}{2}\sum_{\kvec'} {\cal P}_{\kvec\kvec'}
\frac{\Delta_{\kvec'}}{\sqrt{(e_{\kvec'}-\mu)^2 + \Delta_{\kvec'}^2}}\,.
\label{eq:gap}
\end{equation}

The conventional ({\em i.e.\/} ``uncorrelated'' or ``mean-field'') BCS
gap equation \cite{FetterWalecka} is retrieved by replacing the
effective interaction ${\cal P}_{\kvec\kvec'}$ matrix elements by the
matrix elements of the bare interaction.

\subsubsection{Strongly coupled superfluids}
\label{ssec:fullbcs}

The above treatment of a superfluid state has a number of appealing
features. One is that the theory can me mapped onto an ordinary BCS
theory, where the effective interactions and, if applicable, the
single particle spectrum, are calculated for the normal system. The
other is that no assumptions need to be made on the correlation
operator other than that the relevant matrix elements can be
calculated with sufficient accuracy.

The basic assumption of the ``weak coupling'' approximation is that
the superfluid gap at the Fermi surface is small compared to the Fermi
energy. This assumption is not met in low-density neutron matter where
the gap energy can indeed be of the order of half of the Fermi energy;
this is a common feature of practically all neutron matter gap
calculations since the 1970s \cite{YangClarkBCS} until recently
\cite{Gandolfi2009b,ectpaper}. To examine this problem, we have
derived in Ref. \onlinecite{fullbcs} the full variational and
Fermi-HNC theory for a superfluid state of the form \eqref{eq:CBCS}.
Without going into the gory details of this derivation, we mention for
the expert only the central feature: the exchange line
\begin{equation}
  \ell(r\KF) = \frac{\nu}{N}\sum_\kvec n(k) e^{\I\kvec\cdot\rvec}\label{eq:eldef}
  \end{equation}
is replaced by two types of lines,
\begin{equation}
  \ell_v(r) \equiv \frac{\nu}{N}
  \sum_{\kvec} v_{\kvec}^2e^{\I\kvec\cdot\rvec}\qquad\mathrm{and}\qquad
            \ell_u(r) \equiv \frac{\nu}{N}
	   \sum_{\kvec} u_{\kvec}v_{\kvec}e^{\I\kvec\cdot\rvec}
	  \label{eq:luvdef}\,,
\end{equation}
where $\nu=2$ is the degree of degeneracy of the single particle
states, and $N = \nu\sum_\kvec v_{\kvec}^2$. The resulting gap
equation is the same as Eq. \eqref{eq:gap}, the only difference being
that the ingredients ${\cal W}(1,2)$ and ${\cal N}(1,2)$ should be
determined self-consistently for a superfluid system and depend
implicitly on the $\ell_v(r)$ and $\ell_u(r)$.

The second result is more subtle and deserves further discussion: The
Euler equation (\ref{eq:euler}) for a superfluid correlated state
leads to physically incorrect solutions, in fact it has no solutions
for systems that are attractive in the sense that the Landau parameter
$F_0^s < 0$.

\subsection{Analysis of effective interactions}
\label{ssec:effint}

The understanding of the above-mentioned unphysical solutions of the
Euler equations, the construction of effective interactions, and their
choice and consequences for the pairing problem are closely related.
To explain the situation, we must briefly review the relationship
between FHNC-EL theory and parquet diagram summations. We do this for
the simplest case that higher-order exchange diagrams are omitted,
these are quantitatively important even in the low-density limit
\cite{fullbcs}, but do not change the message of our analysis.

The summation of parquet diagrams implies, among others, the summation
of ring diagrams with a local ``particle-hole'' interaction $\tilde
V_{\rm p-h}(q)$. The long wavelength limit is related to the Fermi
liquid parameter
\begin{equation}
  \tilde V_{\rm p-h}(0+) = \frac{2 m}{3m^*}F_0^s\,.\label{eq:F0s}
\end{equation}
One should not expect that this relationship is satisfied exactly
\cite{parquet5} when $\tilde V_{\rm p-h}(0+)$ is obtained by diagram
summations, and $F_0^s$ is from hydrodynamic derivatives; the
{(dis-)agreement} can be taken as a test for the accuracy of the
implementation of the theory \cite{fullbcs,v3eos}.

The density-density response function is in that case
 \begin{equation}
    \chi(q,\omega)= 
    \frac{\chi_0(q,\omega)} {1-\tilde V_{\rm
        p-h}(q)\chi_0(q,\omega)}\label{eq:chiRPA}\,,
  \end{equation}
where $\chi_0(q,\omega)$ is the Lindhard function \cite{Lindhard}. 
The static structure function $S(q)$ is related to the
density--density response function $\chi(q,\omega)$ through
 \begin{equation}
 S(q) = -\int_0^\infty \frac{d\hbar\omega}{\pi} \Im m\chi(q,\omega)\,.
\label{eq:SRPA}
\end{equation}
The connection to the Euler equation (\ref{eq:euler}) of the FHNC-EL
theory is established by assuming a ``collective'' approximation for
the Lindhard function which is constructed such that the $\omega^0$
and $\omega^1$ sum rules are satisfied.
\begin{eqnarray}
        -\Im m\int_0^\infty\frac{ d\hbar\omega}{\pi}\chi_0^{\rm coll}(q,\omega) &&=
        -\Im m\int_0^\infty\frac{ d\hbar\omega}{\pi}\chi_0(q,\omega) = \SF(q)
        \label{eq:m0MSA}\\
        -\Im m\int_0^\infty\frac{ d\hbar\omega}{\pi}\omega\chi_0^{\rm coll}(q,\omega) &&=
        -\Im m\int_0^\infty\frac{ d\hbar\omega}{\pi}\omega\chi_0(k,\omega) = t(q)
\label{eq:m1MSA}
\end{eqnarray}
where $t(q) = \hbar^2 q^2/2m$ and $\SF(q)$ is the static structure function
of the non-interacting Fermi system.
This leads to
\begin{equation}
        \chi_0^{\rm coll}(q,\omega) =
        \displaystyle \frac{2 t(q)}
        { (\hbar\omega+\I\eta)^2-
          \displaystyle{\left(\frac{t(q)}{\SF(q)}\right)^2}}
\label{eq:Chi0Coll}
\end{equation}
and, consequently, to the collective approximation for the
density-density response function
\begin{equation}
        \chi^{\rm coll}(q,\omega) =
        \displaystyle \frac{2 t(q)}
        { (\hbar\omega+\I\eta)^2-
          \displaystyle{\left(\frac{t(q)}{\SF(q)}\right)^2-
            2t(q)\tilde V_{\rm p-h}(q)}} \,.
\label{eq:ChiColl}
\end{equation}
In this case, the frequency integration \eqref{eq:SRPA} can be carried
out analytically, which leads to the simplest form of the Euler
equation of FHNC-EL theory \cite{polish}
\begin{equation}
  S(q) = \frac{\SF(q)}{\sqrt{1+\displaystyle\frac{2 \SF^2(q)}{t(q)}\tilde V_{\rm p-h}(q)}}\,.
  \label{eq:Scoll}
\end{equation}
Since $S(q)\propto q$ for $q\rightarrow 0+$, negative values of
$\tilde V_{\rm p-h}(q)$ and, hence, negative values of $F_0^s$ are
permitted.

In the superfluid system, the variational principle \eqref{eq:euler}
leads to the same equation \eqref{eq:Scoll}, a small additional term
\cite{fullbcs} does not change our analysis. However, the
static structure function has the form
\begin{equation}
  \SF(q) = 1 - \frac{\rho}{\nu}
	  \int d^3 r e^{\I \qvec\cdot\rvec}\left[
	    \ell_v^2(r) - \ell_u^2(r)\right]
	\,.\label{eq:SFdef}
\end{equation}
It follows immediately from the definitions \eqref{eq:luvdef} that the
long-wavelength limit of $\SF(q)$ is
\begin{equation}
  \SF (0+) = 2\frac{\sum_{\kvec} u_{\kvec}^2 v_{\kvec}^2}{\sum_{\kvec}v_{\kvec}^2} > 0\,.
\label{eq:SF0}
\end{equation}
Hence, $\SF(0+) > 0$ for the superfluid system. As a consequence,
Eq. \eqref{eq:Scoll} has no sensible solution of $F_0^s < 0$ even for
an infinitesially small but finite gap.

The problem is readily solved by abandoning the ``collective''
approximation \eqref{eq:ChiColl}, in other words moving from the pure
Jastrow-Feenberg wave function to the parquet summations. There have
been several suggestions for a Lindhard function for a superfluid
system
\cite{PhysRevB.61.9095,PhysRevA.74.042717,Steiner2009,Vitali2017}, the
most frequently used form for $T=0$ is given below. In the superfluid
case, $\chi_0(q,\omega)$ also depends on the spins. In terms of the
usual relationships of BCS theory,
\begin{eqnarray}
  u_k^2 &=& \frac{1}{2}\left(1+\frac{\xi_\kvec}{E_\kvec}\right)
  \nonumber\\
  v_k^2 &=& \frac{1}{2}\left(1-\frac{\xi_\kvec}{E_\kvec}\right)\,.
\end{eqnarray}
with $\xi_{\kvec} = t(k)-\mu$ and $E_{\kvec} =
\sqrt{\xi_{\kvec}^2+\Delta_{\kvec}^2}$
we have \cite{Schrieffer1999,Kee1998,Kee1999,PhysRevB.61.9095}

\begin{equation}
  \chi_0^{(\rho,\sigma)}(\kvec,\omega) = \frac{\nu}{N}\sum_{\pvec}
  b_{\pvec,\kvec}^{(\rho,\sigma)}\Biggl[\frac{1}
    {\hbar\omega-E_{\kvec+\pvec}-E_{\pvec}+\I\eta} -
    \frac{1}{\hbar\omega+E_{\kvec+\pvec}+E_{\pvec}+\I\eta}\Biggr]
  \label{eq:BCSLindha}
\end{equation}
with
\begin{eqnarray}
  b_{\pvec,\kvec}^{(\rho,\sigma)} &=&\frac{1}{4}\left[\left(1-
    \frac{\xi_{\pvec}}{E_{\pvec}}\right)
    \left(1+\frac{\xi_{\kvec+\pvec}}{E_{\kvec+\pvec}} \right)
    \pm\frac{\Delta_{\pvec}}{E_{\pvec}}
    \frac{\Delta_{\kvec+\pvec}}{E_{\kvec+\pvec}}\right]
  \nonumber\\ &=&v_{\pvec}^2u_{\kvec+\pvec}^2 \pm u_{\pvec} v_{\pvec}
  u_{\kvec+\pvec} v_{\kvec+\pvec}\,,
\end{eqnarray}
where the upper sign applies to the density channel, and the lower
to the spin channel, respectively.

A similar analysis applies to the effective interaction ${\cal W}(1,2)$
and the energy numerator term  ${\cal N}(1,2)$.
In principle, these two quantities are non-local $2$-body
operators. The leading, local contributions to these operators are
readily expressed in terms of the diagrammatic quantities of FHNC-EL
theory \cite{polish}:
\begin{eqnarray}
{\cal N}(1,2) &=& {\cal N}(r_{12})\, =\, \Gamma_{\rm dd}(r_{12})\,,\nonumber\\
{\cal W}(1,2) &=& W(r_{12})\,,
\label{eq:NWloc}
\end{eqnarray}
where $\Gamma_{\rm dd}(r_{12})$ is the ``direct correlation function''
of FHNC theory \cite{Johnreview,polish}.  In an approximation
corresponding to the one spelled out in Eqs. \eqref{eq:Scoll} we have
\begin{eqnarray}
  \tilde\Gamma_{\rm dd}(q) &=& \frac{1}{\SF(q)}
  \left[\left[1+\displaystyle\frac{2 \SF^2(q)}{t(q)}\tilde V_{\rm p-h}(q)\right]^{-1/2}-1\right]\label{eq:gdddef}\\
  \widetildeto{\Gamma}{W}(q) &=&  \frac{t(q)}{\SF^2(q)}
  \left[1-\left[1+\displaystyle\frac{2 \SF^2(q)}{t(q)}\tilde V_{\rm p-h}(q)\right]^{-1/2}\right]= -\frac{t(q)}{\SF(q)}\tilde
  \Gamma_{\rm dd}(q)\label{eq:Wdef}\,.
\end{eqnarray}
These relationships display the same problems as the $S(q)$ above,
namely that they lead to unphysical results for negative $F_0^s$. The
solution is again found by examining the construction of
$ \widetildeto{\Gamma}{W}(q)$ from the viewpoint of perturbation theory.

Eq.~(\ref{eq:chiRPA}) defines an {\em energy
  dependent\/} effective interaction $\widetildeto{\Gamma}{W}(q,\omega)$
which we write as the sum of the energy independent
term $\tilde V_{\rm p-h}(q)$ and the energy dependent induced interaction
$\widetildeto{\Gamma}{W}_{\rm I}(q,\omega)$
 \begin{equation}
   \widetildeto{\Gamma}{W}(q,\omega) = \frac{\tilde V_{\rm p-h}(q)} {1-\tilde V_{\rm
       p-h}(q)\chi_0(q,\omega)} = \tilde V_{\rm p-h}(q) +
   \frac{\tilde V_{\rm p-h}^2(q)\chi_0(q,\omega)} {1-\tilde V_{\rm
       p-h}(q)\chi_0(q,\omega)}\,.
   \label{eq:Wnonlocal}
 \end{equation}
 An {\em energy independent\/} effective interaction
 $\widetildeto{\Gamma}{W}(q)$ is then
defined such that it leads to the same $S(q)$, {\ie \/}
\begin{eqnarray}
S(q) &=& -\int_0^\infty \frac{d\hbar\omega}{\pi} \Im m
 \frac{\chi_0(q,\omega)} {1-\tilde V_{\rm
     p-h}(q)\chi_0(q,\omega)}\nonumber\\
 &=& -\int_0^\infty \frac{d\hbar\omega}{\pi} \Im m
 \left[\chi_0(q,\omega)+ \chi^2_0(q,\omega) \widetildeto{\Gamma}{W}(q,\omega)\right]
 \nonumber\\
  &\overset{!}{=}& 
-\int_0^\infty \frac{d\hbar\omega}{\pi} \Im m
\left[\chi_0(q,\omega)+ \chi^2_0(q,\omega) \widetildeto{\Gamma}{W}(q)\right]\,,
\label{eq:Wlocal}
\end{eqnarray}
where the last line defines $\widetildeto{\Gamma}{W}(q)$ and, through
Eq. \eqref{eq:Wnonlocal}, the static induced interaction
$\widetildeto{\Gamma}{W}_{\rm I}(q) =\widetildeto{\Gamma}{W}(q)-\tilde V_{\rm
     p-h}(q) $\,.
If we furthermore use
the collective approximation \eqref{eq:ChiColl} for
$\chi_0(q,\omega)$, Eq. \eqref{eq:Wdef} follows.

Realizing these connections there is, of course, no reason for not
using the full Lindhard functions for defining the effective
interaction $\widetildeto{\Gamma}{W}(q)$ in \eqref{eq:Wlocal}. This
can be done using the Lindhard function for the normal system, or
\eqref{eq:BCSLindha}. The latter is numerically rather demanding, we
have carried this out in Ref. \onlinecite{fullbcs}. It turns out that
the use of \eqref{eq:BCSLindha} makes little difference for $\tilde
V_{\rm p-h}(q)$, we have therefore used in our ground state
calculations the Lindhard function for the normal system.

This is different for the gap equation, partly due to the exponential
dependence of the superfluid gap on the interaction strength.  We have
therfore used \eqref{eq:BCSLindha} and $\omega=0$ for the effective
interactions in the pairing calculation which is more appropriate for
these low-energy phenomena \cite{SPR2001}.

\subsection{Analysis of the gap equation}
\label{ssec:gapeq}

The appearance of the ``energy numerator'' term in the pairing
interaction matrix element \eqref{eq:Pdef} is a feature that might be
unfamiliar to the reader who is only familiar with mean-field
theories, but it comes in quite naturally when the gap equation is
expressed in terms of the $T$-matrix \cite{PethickSmith}.  This
section is devoted to a discussion of the importance of this term
which arises in an expansion of the correlated BCS state
\eqref{eq:CBCS} in the number of Cooper pairs. We stress again that no
assumption on the nature of the correlation operator has been made in
the derivation.

If the gap at the Fermi surface is small, we can replace the pairing
interaction $\tilde{\cal W}(k)$ by its $S$-wave matrix element at the
Fermi surface,
\begin{equation}
\tilde {\cal W}_F \equiv \frac{1}{2 \KF^2}\int_0^{2\KF} k dk \widetildeto{\Gamma}{W}(k)
= N{\cal W}_{\KF,\KF}\,.
\label{eq:V1S0}
\end{equation}
Then we can write the gap equation as
\begin{equation}
1 = - \tilde {\cal W}_F\int\frac{ d^3k'}{(2\pi)^3\rho}
\Bigg[\frac{1}{\sqrt{(e_{k'}-\mu)^2 + \Delta^2_{\KF}}}\label{eq:gaplowdens}
  -\frac{|e_{k'}-\mu|}{\sqrt{(e_{k'}-\mu)^2 + \Delta^2_{\KF}}}
\frac{\SF(k')}{t(k')} \Bigg]\,,
\end{equation}
which is almost identical to Eq.~(16.91) in
Ref.~\citenum{PethickSmith}.  In particular, the second term, which
originates from the energy numerator generated in Eq.~(\ref{eq:gap})
by the second term of ${\cal P}_{\kvec\kvec'}$ in Eq.~(\ref{eq:Pdef}),
has the function of regularizing the integral for large $k'$.

This observation leads us to two conclusions: First, the effective
interaction $\widetildeto{\Gamma}{W}(k)$ should be identified with a
local approximation to the $T$-matrix. This is also evident because
its diagrammatic structure contains both particle-particle and
particle-hole reducible diagrams. Second, a correct balance between
the energy numerator and the interaction term are essential to
guarantee the convergence of the integral.

When the gap is large, one can no longer argue that the energy
numerator, which vanishes at the Fermi surface, is negligible.  The
convergence of the integrals is, in this case, guaranteed by the fact
that the interactions fall off for large $k'$. Since the integrals
would diverge if the interactions did not fall off, the precise
asymptotic form can have a profound quantitalive influence on the
magnitude of the gap.

To be more precise, we can again study the behavior of
the integrand for large $k'$:
\begin{eqnarray}
{\cal P}_{\kvec\kvec'} &=& {\cal W}_{\kvec\kvec'}+(|e_{\kvec}- \mu | 
+ |e_{\kvec'}- \mu |){\cal N}_{\kvec\kvec'}
\nonumber\\
&\rightarrow& {\cal W}_{0,\kvec'}+ t(k'){\cal N}_{0,\kvec'}
\end{eqnarray}
From Eq. \eqref{eq:Wdef} we can now conclude that these two terms always
cancel for large arguments.

The cancellation of these two terms is, of course, a consequence of
either the functional optimization of the correlations, or the parquet
diagram summations. It is therefore expected that the actual value of
the gap depends sensitively on how the energy numerator is treated.
This also applies to the question of how one should deal with a
non-trivial single particle spectrum; comments on this are found in
Ref. \onlinecite{ectpaper}. Similar concerns apply to calculations
that use state-independent correlation functions of the form
\eqref{eq:Jastrow}, including our own work \cite{ectpaper}: The
correlations are optimized for the central channel of the interaction,
but the paring interaction is calculated in the
singlet-$S$-channel. Hence, the cancellations between energy numerator
and interaction term are violated. The alternative, namely calculating
the correlations for a model where the singlet-$S$ channel is taking
as state-independent interaction is not a viable one because such a
system would become unstable against infinitesimal density
fluctuations at densities much smaller than those of interest here.

Finally, we go back to the seminal paper by Cooper, Mills, and Sessler
\cite{CMS} who showed that the gap equation has indeed solutions for
interactions with strongly repulsive cores. Taming the strongly
repulsive core of the nucleon-nucleon interaction was also the
original intention of the Jastrow method \cite{Jastrow55}, one might
therefore legitimately ask if using Jastrow correlation in combination
with a BCS state does not double count the short-ranged correlations.

As long as the theory is based on a clean expansion in the number of
Cooper pairs, there is by construction no overcounting problem, but it
is instructional to see the interplay between Jastrow-correlations and
BCS correlations. To see that, it is sufficient to examine the
two-body approximation which is still occasionally being used
\cite{Fabrocinipairing,Fabrocinipairing2,Pavlou2017,Benhar}.  We also
restrict ourselves, for simplicity, to state-independent correlations.
In that approximation, we have
\begin{equation}
  W^{(2)}(r) = f^2(r) v(r) + \frac{\hbar^2}{m}\left|\nabla f(r)\right|^2\,,
  \qquad \Gamma_{\rm dd}^{(2)}(r) = f^2(r)-1 \equiv h(r)\label{eq:2body}\,.
\end{equation}
Following Ref. \onlinecite{CMS}, Eq. (15), we introduce
\begin{equation}
  \tilde\chi(k) =\frac{1}{2}
  \frac{\Delta_{\kvec'}}{\sqrt{(e_{\kvec'}-\mu)^2 + \Delta_{\kvec'}^2}}\,.
  \label{eq:chidef}
  \end{equation}
The short-ranged structure of the correlations is determined by the
short-wavelength behavior of the gap equation, in that case we get for
the coordinate space representation of the right-hand side of the gap equation
as
\begin{eqnarray}
  &&\left[-\frac{\hbar^2}{2m}\nabla^2-\mu\right]h(r)\chi(r)
  + h(r)\left[-\frac{\hbar^2}{2m}\nabla^2-\mu\right]\chi(r)
  + \left[f^2(r)v(r)+\frac{\hbar^2}{m}\left|\nabla f(r)\right|^2\right]
  \chi(r)\nonumber\\&=&
  f(r)\chi(r)\left[-\frac{\hbar^2}{m}\nabla^2 + v(r)\right]f(r)
  -\frac{\hbar^2}{m}\nabla\cdot(h(r)\nabla \chi(r))-2\mu h(r)\chi(r)\,.
  \label{eq:BCS2ndorder}
  \end{eqnarray}
Above, the first two terms come from the energy numerator and the last
from the interaction. If we assume that the correlation function is
determined by a Schr\"odinger-like equation as, for example, in the
LOCV method, the Jastrow correlation function serves to cancel the
short-ranged interaction. Combining these terms as in the second line
shows how the Jastrow correlation function $f(r)$ eliminates the
short-ranged part of the interaction, leaving $\chi(r)$ to deal with
BCS-specific correlations. On the other hand, ignoring the energy
numerator term destroys this cancellation.

A further evidence for the delicate balance between the two terms
in the pairing interaction \eqref{eq:Pdef} is uncovered by
calculating the particle-hole average
\[\sum_{ph}\left[{\cal W}_{\pvec\hvec'}+(|e_{\pvec}- \mu | 
+ |e_{\hvec}- \mu |) {\cal N}_{\pvec\hvec}\right]=
\sum_{ph}\left[{\cal W}_{\pvec\hvec}+(e_{\pvec}- e_{\hvec}) {\cal
    N}_{\pvec\hvec}\right] \,.\]
Using a free single-particle spectrum and the local approximations
\eqref{eq:gdddef} and \eqref{eq:Wdef}, we find that this average is
zero. This is actually only a special case of a more general statement
that the particle-hole average of CBF effective interaction is zero
for optimized correlation functions.

This means, of course, that both the ``average zero'' property and the
cancellation of the short-ranged structure of the interaction does not
apply for cases where the correlation functions are optimized for,
say, the central part of the interaction, but then the singlet
projection is used for the pairing calculation.

\section{Application to Neutron Matter}
\label{sec:results}

\subsection{General Remarks}
\label{ssec:energetics}

We have carried out calculations for static properties and superfluid
pairing gaps in neutron matter based on two representative NN
interactions acting in the $T=1$ channel, namely the $v_6$ version of
the Reid soft-core potential \cite{Reid68} as formulated in
Eqs.~(A.1)-(A.8) of Ref.~\citenum{Day81}, and the Argonne $v_6$
potential \cite{AV18}. Several types of calculations were done:
Parquet calculations as described in Ref.~\citenum{v3eos}, and parquet
calculations including the most important non-parquet corrections, the
so-called ``twisted chain'' diagrams \cite{v3twist}. The calculations
for the ground state calculations were all done for the normal
system. We have, in Ref. \onlinecite{fullbcs}, also used the
superfluid Lindhard functions \eqref{eq:BCSLindha} which requires a
rather demanding numerical calculation to capture the sharp structures
of the integrands around the Fermi surface.  In that work, we have
determined that this causes no visible change in the essential inputs
for the pairing interaction, even if the gap is of the order of half
the Fermi energy.

For the calculation of the effective interactions, we have used both
the normal Lindhard function as well as the generalizations
\eqref{eq:BCSLindha} to superfluid systems.

\subsection{Effective interactions}

Let us return to the effective interactions \eqref{eq:Wdef}.
Following the discussions of sections \ref{ssec:effint} and
\ref{ssec:gapeq}, we can write the induced interaction in the
state-dependent parquet scheme as
   \begin{equation}
     \widetildeto{\Gamma}{W}_{\rm I}^{(\alpha)}(q,0) =
     \frac{\left[\tilde V_{\rm p-h}^{(\alpha)}(q)\right]^2
       \chi_0^{(\alpha)}(q,0)} {1-\tilde V_{\rm
         p-h}^{(\alpha)}(q)\chi_0^{(\alpha)}(q,0)}\,.
   \label{eq:Walphanonlocal}
   \end{equation}
where the superscript $\alpha$ refers to the operator channel $\1$, $\hat
L \equiv ({\bm\sigma}_1\cdot \hat\rvec)({\bm\sigma}_2\cdot \hat\rvec)$,
and $\hat T 
\equiv {\bm\sigma}_1\cdot{\bm\sigma}_2- ({\bm\sigma}_1\cdot
\hat\rvec)({\bm\sigma}_2\cdot\hat\rvec)$.

We have included in the induced interaction
$\widetildeto{\Gamma}{W}_{\rm I}(q) \equiv
\widetildeto{\Gamma}{W}_{\rm I}(q,\omega=0)$ the leading exchange
diagram which are important to establish a reasonable agreement
between the long-wavelength limit of the particle-hole interaction and
Landau's Fermi-liquid parameter $F_0^s$, \ie we use for the
particle--hole interaction in Eq. \eqref{eq:Wnonlocal} in the $\{\1,
\hat L, \hat T\}$ channel basis
\begin{equation}
  \tilde V_{\rm p-h}^{(\alpha)}(q)= \tilde V_{\rm p-h,d}^{(\alpha)}(q)
  + \tilde V_{\rm p-h,ex}^{(\alpha)}(q)\,,
\end{equation}
where the $\tilde V_{\rm p-h,ex}^{(\alpha)}(q)$ is calculated as
spelled out in the Appendix of Ref. \onlinecite{v3eos}, and
$\chi_0^{(L)}(q,0)=\chi_0^{(T)}(q,0)\equiv\chi_0^{(\sigma)}(q,0)$.
One can go beyond this relatively simple approximation and include
higher-order exchange diagrams, these would, among others, establish
the correct relationships between the sum rules for the Fermi Liquid
parameters and those for the forward scattering amplitudes
\cite{BabuBrown}. The effect may be important at higher densities and
in the case of $P$-wave pairing \cite{PhysRevC.99.014310}. However,
the most important input to the calculation is the particle-hole
irreducible interaction $\hat V_{\rm p-h}(q)$. This should {\em not\/}
be identified with some local approximation of the $G$-matrix. This is
seen most easily in a self-bound system like nuclear matter by the
simple argument that the Fermi-sea average of the $G$ matrix should
basically be the interaction correction to the binding energy which is
negative. On the other hand, the matrix element of $V_{p-h}^{(\1)}(r)$
at the Fermi surface is the interaction correction to the
incompressibility which is positive \cite{Shlomo06}. The more
important consideration is, in our opinion, to establish a reasonably
accurate agreement between the long-wavelength limit (\ref{eq:F0s})
and the hydrodynamic compressinility
\begin{equation}
mc^2 = \frac{d}{d\rho}\rho^2 \frac{d}{d\rho}\frac{E}{N}
 = mc_{\rm F}^{*2} + \tilde V_{\rm p-h}(0+) \equiv  mc_{\rm F}^{*2}(1+F_0^S)
\,,
\label{eq:FermimcfromVph}
\end{equation}
where $c_{\rm F}^* = \sqrt{\frac{\hbar^2\KF^2}{3mm^*}}$ is the speed
of sound of the non-interacting Fermi gas with the effective mass
$m^*$.  We have discussed this issue in Ref. \onlinecite{v3eos}.

Our work goes beyond previous calculations in two important
aspects. One is the full execution of the localized parquet diagrams,
including the ``twisted chain'' diagrams that go beyond the parquet
class. The second is the use of a Lindhard function
\eqref{eq:BCSLindha} appropriate for superfluid systems. Both of these
corrections are expected to be most visible at low densities, but for
different reasons:

The bare singlet interaction is close to forming a bound state;
therefore a small change in the effective interaction can cause a
rather large change in the short-ranged correlations \cite{v3lett}.  A
very careful evaluation of all relevant quantities is therefore
essential.

Moreover, at low densities, the superfluid gap is about half of the
Fermi energy, therefore there is no reason to assume that the use of a
Lindhard function appropriate for a normal system is justified. Note
also that $\lim_{q\rightarrow 0}\chi^{(\sigma)}(q,0) = 0$, \ie the use
of a superfluid Lindhard function suppresses the induced interactions
$\widetildeto{\Gamma}{W}^{(L)}(q)$ and
$\widetildeto{\Gamma}{W}^{(T)}(q)$ in the long wavelength limit.

Let us therefore go through the individual steps. All calculations
refer to the $v_6$ version of the Argonne potential \cite{AV18}, we
have chosen a density of $\KF = 0.4\,\mathrm{fm}^{-1}$ where the
superfluid gap is close to its maximum value as a function of density,
and to $\KF = 0.8\,\mathrm{fm}^{-1}$ where it is declining but still
visible.  Input to the calculations are the particle-hole irreducible
interactions $\tilde V_{\rm p-h}^{(\alpha)}(q)$ and the Lindhard
functions $\chi_0^{(\alpha)}(q,0)$. We show these for the above two
typical values of $\KF$ in Figs.  \ref{fig:Vphplot04} and
\ref{fig:Vphplot08}.
\begin{widetext}
  \begin{figure}
  \includegraphics[width=0.34\textwidth,angle=270]{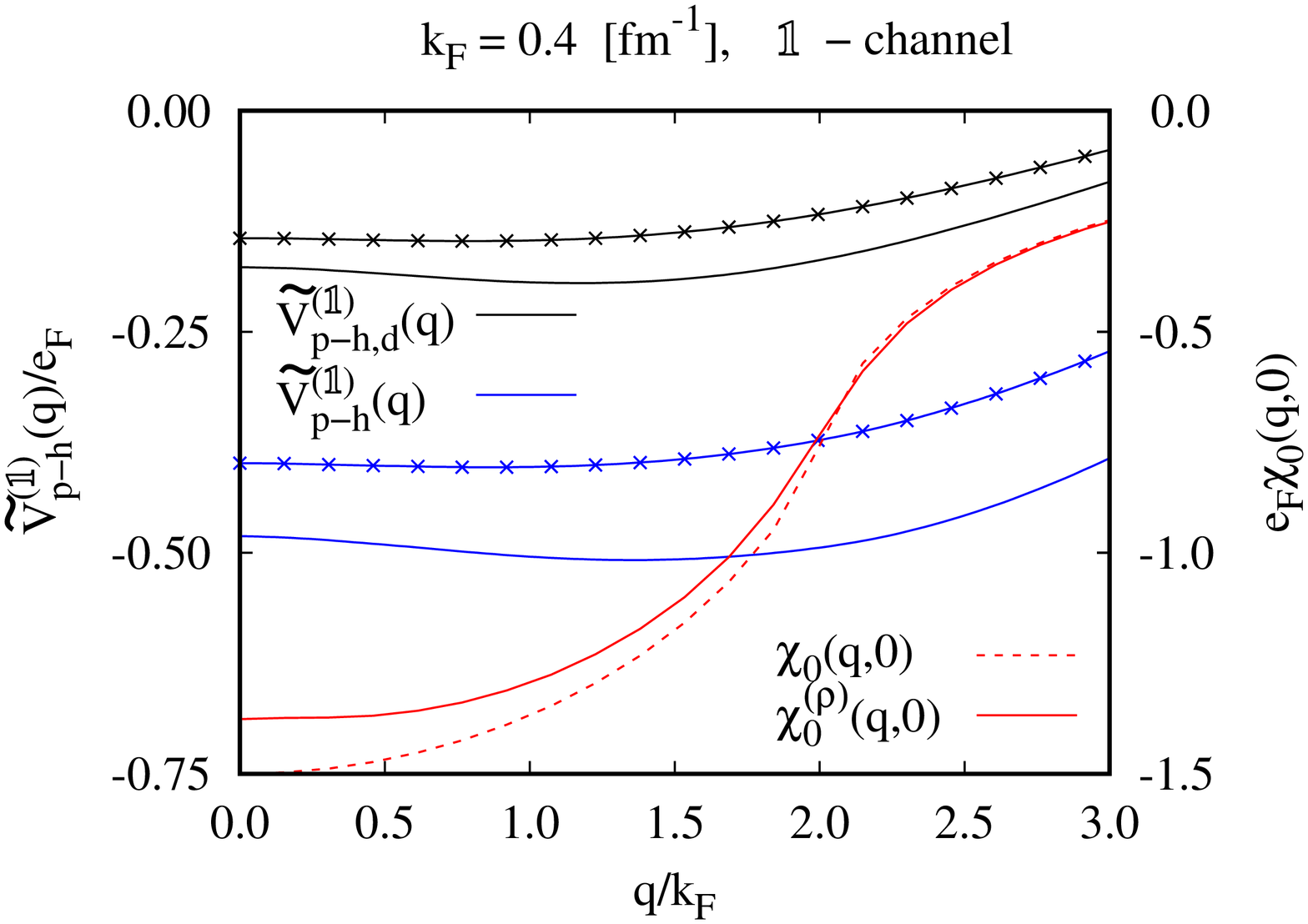}
    \includegraphics[width=0.34\textwidth,angle=270]{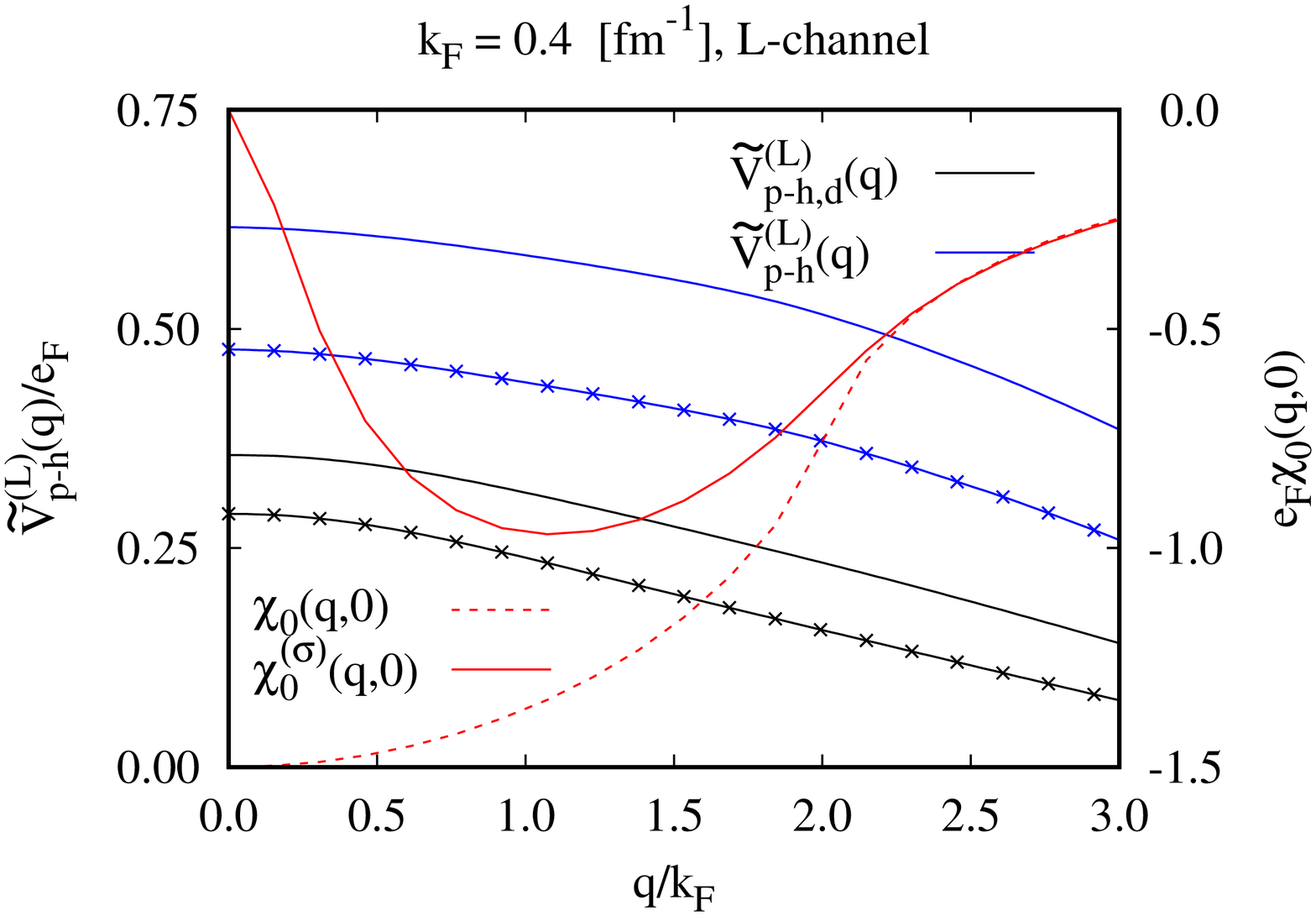}
    \caption{The figures show, for $\KF = 0.4\,\mathrm{fm}^{-1}$, the
      central and longitudinal components of both the ``direct''
      particle-hole interaction $\tilde V_{\rm p-h,d}^{(\alpha)}(q)$
      (black lines, left scale), and effective interaction including
      exchange diagrams $\tilde V_{\rm p-h}^{(\alpha)}(q)\equiv \tilde
      V_{\rm p-h,d}^{(\alpha)}(q)+\tilde V_{\rm p-h,ex}^{(\alpha)}(q)$
      (blue lines, left scale.) We show both the parquet results and
      those including non-parquet corrections (lines with markers).
      Also shown are the normal system Lindhard function (red dashed
      line, right scale) and the superfluid system Lindhard function
      (red solid line) in the density (left panel) and spin channel
      (right panel), respectively.\label{fig:Vphplot04}}
\end{figure}
  \begin{figure}
  \includegraphics[width=0.34\textwidth,angle=270]{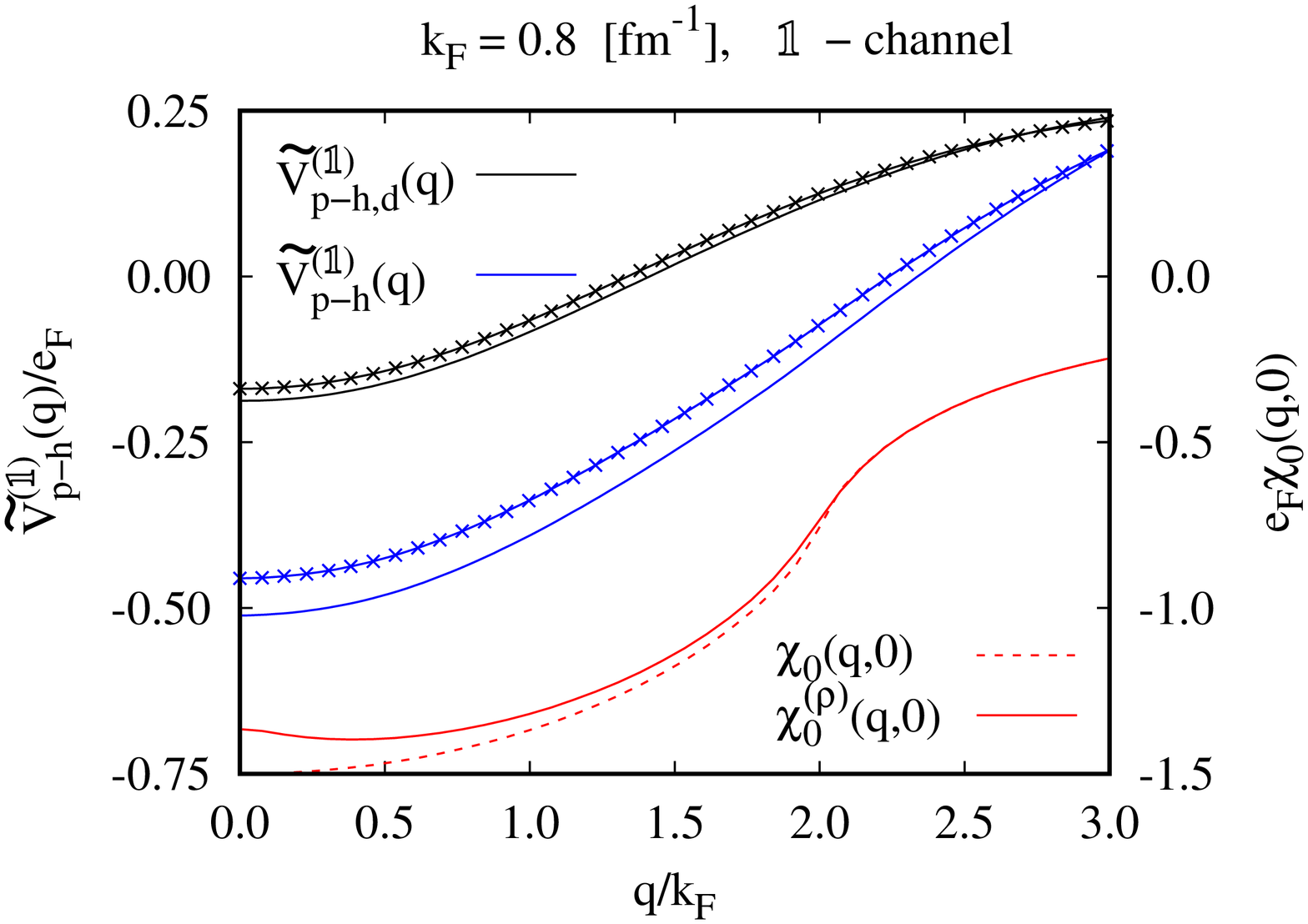}
    \includegraphics[width=0.34\textwidth,angle=270]{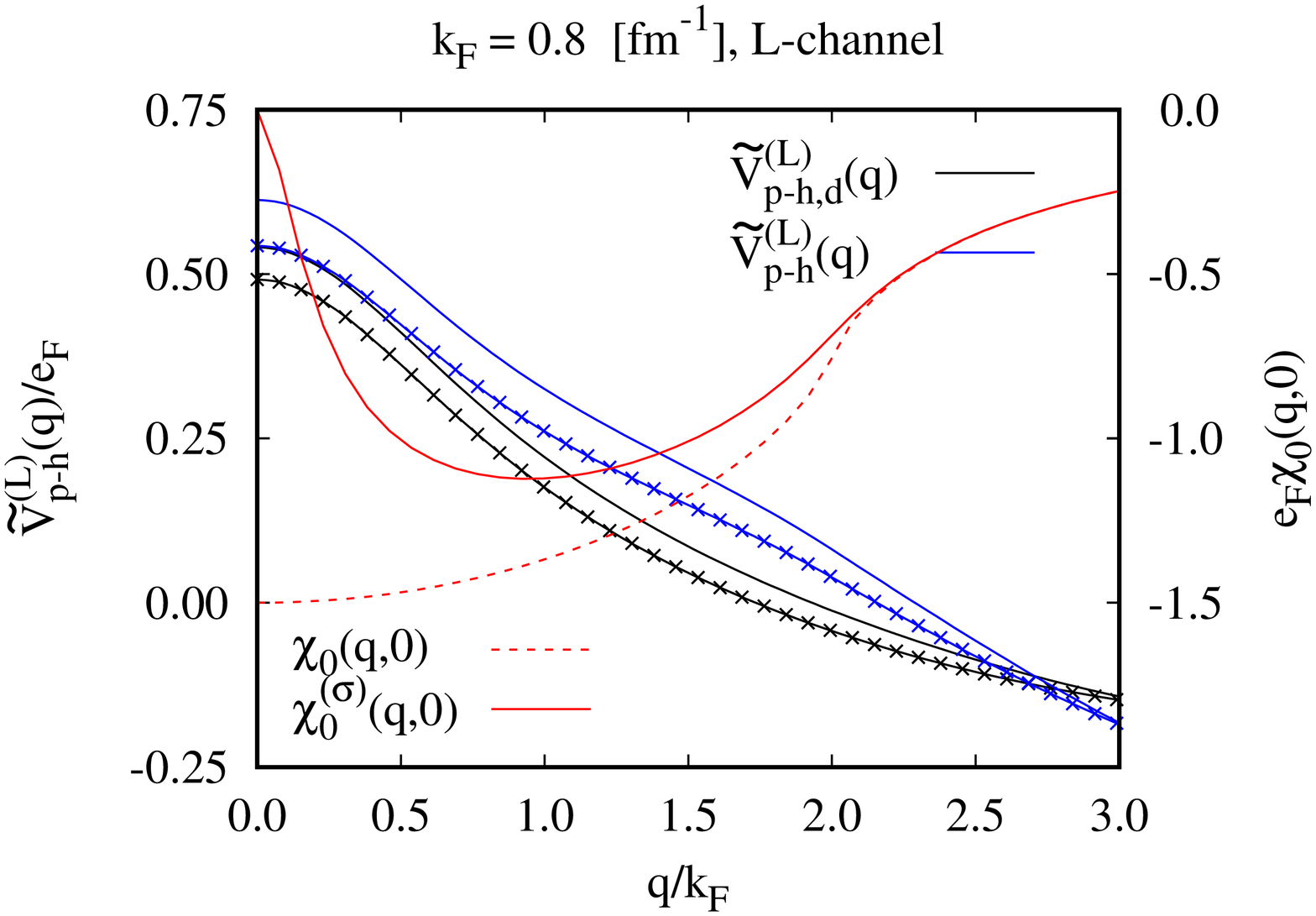}
    \caption{Same as Fig. \ref{fig:Vphplot04} for
      $\KF = 0.8\,\mathrm{fm}^{-1}$.\label{fig:Vphplot08}
    }
  \end{figure}
\end{widetext}

Practically all of these results look rather innocuous. As shown in
our previous work \cite{v3eos}, the inclusion of exchange diagrams is
important to have a reasonably accurate relationship between the Fermi
liquid parameters obtained from hydrodynamic derivatives and the
long-wavelength limit of the particle-hole interaction. The ``twisted
chain'' diagrams are the most pronounced many-body correction at low
densities \cite{v3twist}, but their effect is moderate. Considering
the exponential dependence of the superfluid gap on the interaction
strength, these processes can, of course, be quantitatively relevant.

The superfluid Lindhard function deviates, in the density channel, by
about 10 to 20 percent from the normal system Lindhard function. The
most pronounced new effect is that the superfluid Lindhard function in
the spin-channel, $\chi_0^{(\sigma)}(q,0)$, goes to zero in the long
wavelength limit.  At low densities, $\KF = 0.4\,\mathrm{fm}^{-1}$,
this falloff already happens at $q=\KF$ which has the effect of
suppressing the induced interaction. As expected, all corrections
become smaller with increasing density. In the case of the superfluid
Lindhard function, this is partly the case due to the smaller value of
the gap, but evidently the correction in the spin channel is still
quite visible.

Turning to the interactions that actually go into the gap equation, we
show in Figs. \ref{fig:twistplot04} and \ref{fig:twistplot08} the
interaction $\widetildeto{\Gamma}{W}(q)$ appearing in
Eq. \eqref{eq:Wnldef}. A somewhat surprising, but easily understood,
feature is the rather dramatic consequence of using the superfluid
Lindhard function in the density channel: The fact that the
long-wavelength limit $\tilde V_{\rm p-h}(0+)$ is of the order of
$-0.5\EF$, and that value of the Lindhard function changes by about 20
percent, can change the induced interaction by a factor of 2 which is
seen in the left part of Fig. \ref{fig:twistplot04}. This finding is
consistent with the observation that the effect is smaller when
non-parquet diagrams are included because the magnitude of $\tilde
V_{\rm p-h}(0+)$ is decreased. Of course, it must be kept in mind that
the agreement between the $F_0^s$ obtained from $\tilde V_{\rm
  p-h}(0+)$, see Eq. \eqref{eq:F0s}, and that obtained from the
hydrodynamic speed of sound, Eq. \eqref{eq:FermimcfromVph}, is only
approximate \cite{v3eos}.

The similarly significant change of the longitudinal part of the
effective interaction, as shown in the right part of Fig.
\ref{fig:twistplot04}, is much more expected and comparable in both
parquet and ``beyond parquet'' results. As we go to higher density,
see Figs. \ref{fig:twistplot08}, the effects become smaller simply due
to the fact that the superfluid gap becomes smaller, but they are
still quite visible.
\begin{widetext}
  \begin{figure}
  \includegraphics[width=0.34\textwidth,angle=270]{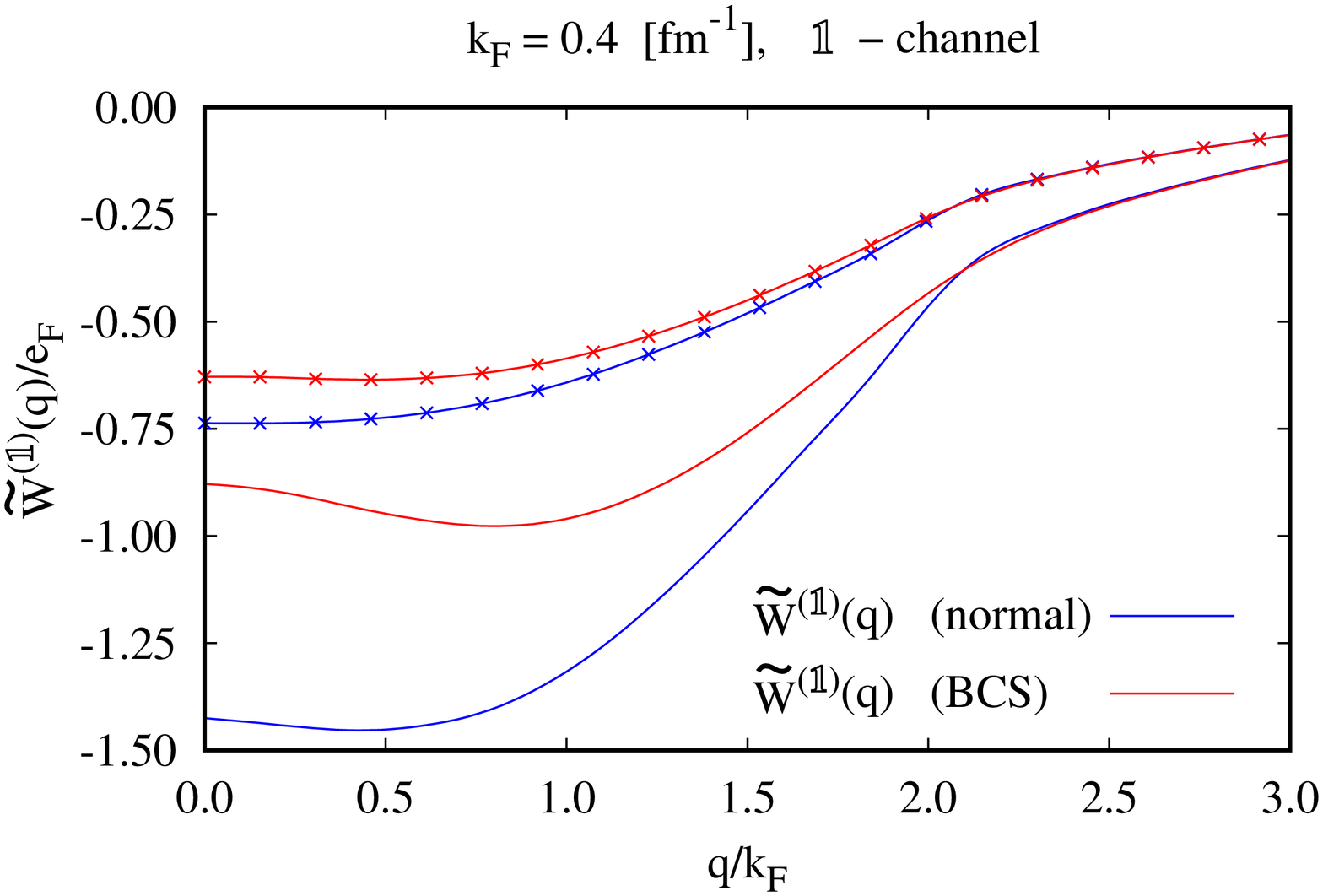}
  \includegraphics[width=0.34\textwidth,angle=270]{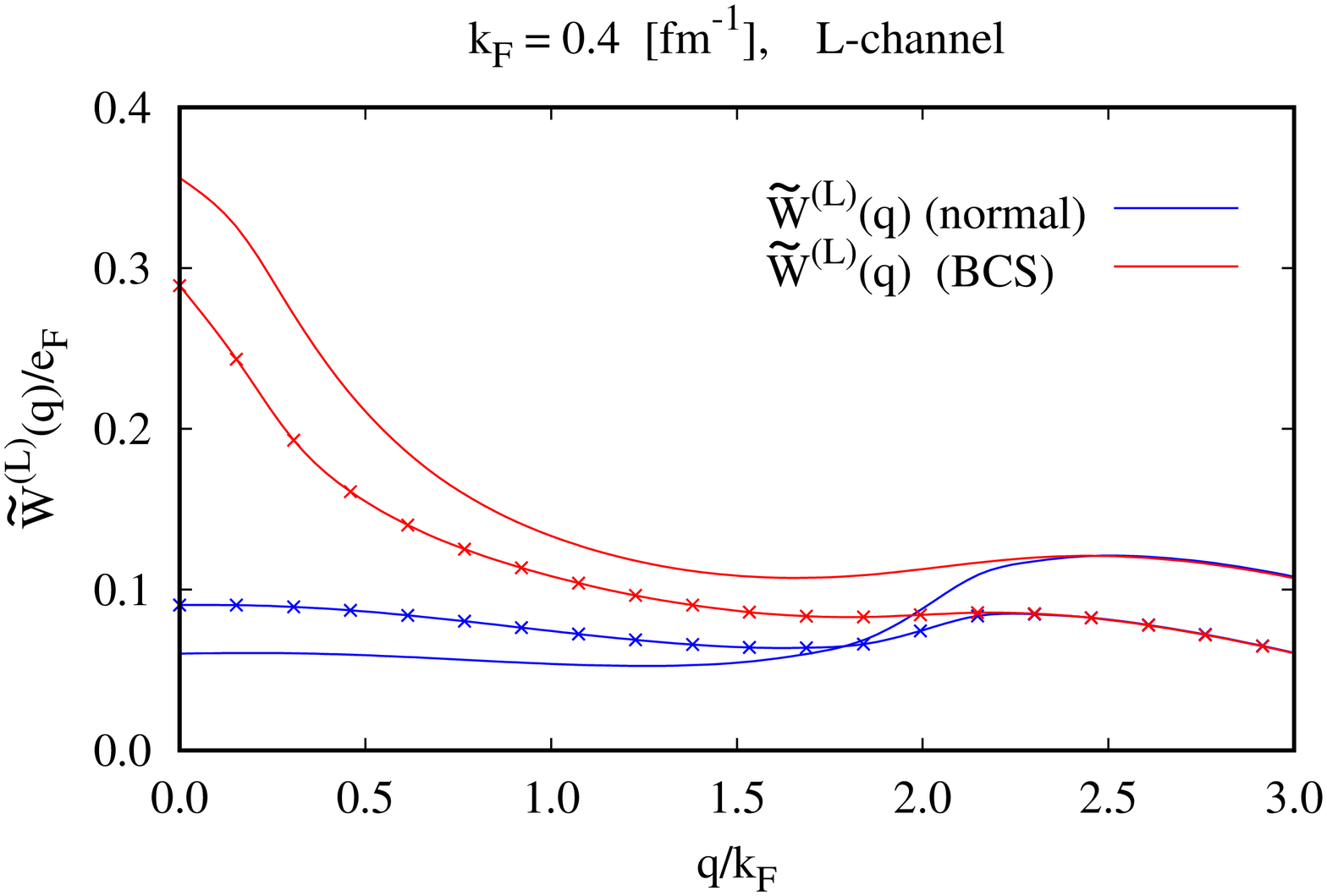}
   \caption{The figures shows, for $\KF = 0.4\,\mathrm{fm}^{-1}$, the
 central (left figure) and longitudinal components (right figure)
     of the effective interactions
     $\widetilde{W}^{(\alpha)}(q)$,
     using the normal system
      Lindhard functions (blue lines) and the superfluid system
      Lindhard functions (red lines). The results including ``beyond
      parquet'' diagrams are marked with
      crosses.\label{fig:twistplot04}}
\end{figure}

  \begin{figure}
  \includegraphics[width=0.34\textwidth,angle=270]{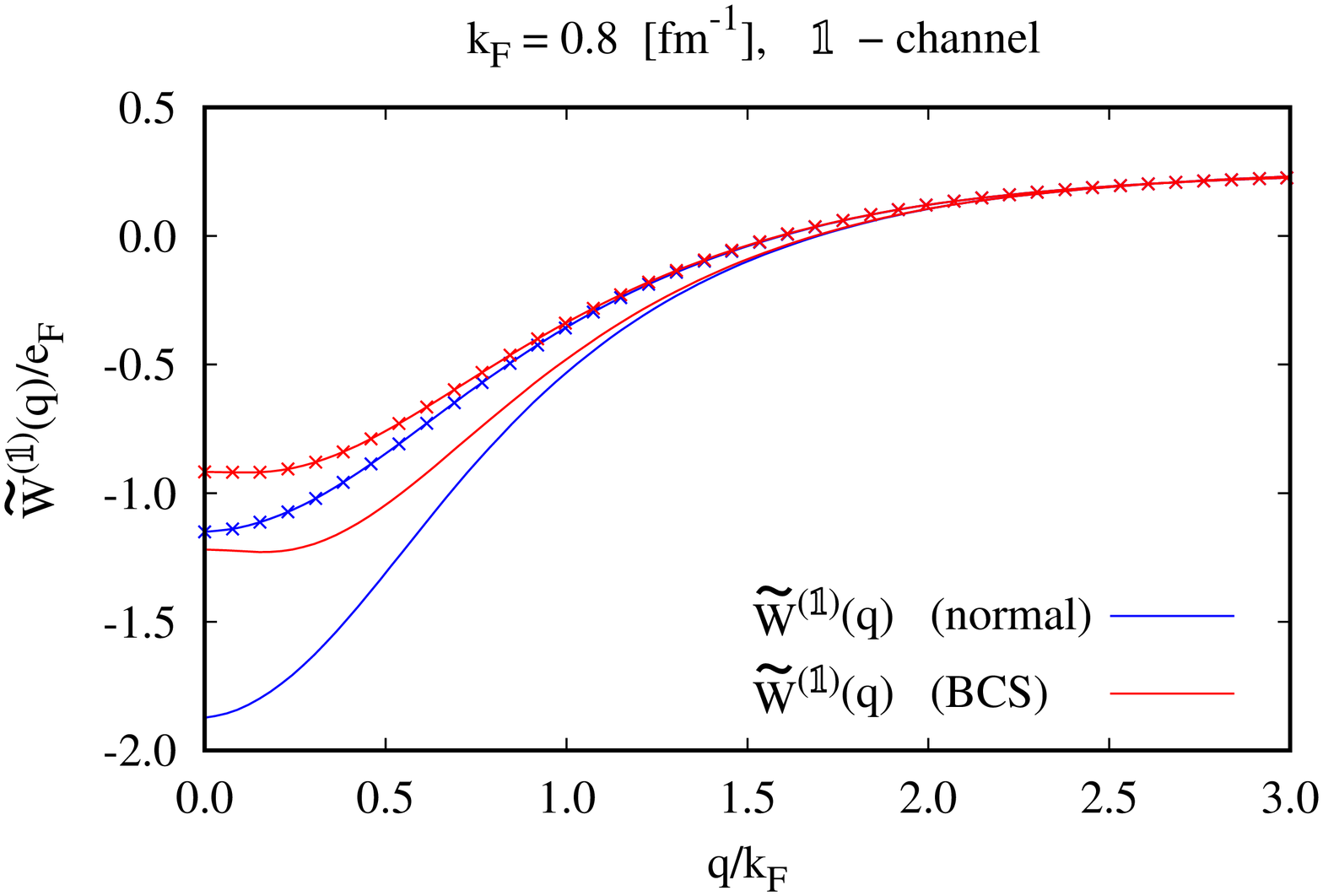}
    \includegraphics[width=0.34\textwidth,angle=270]{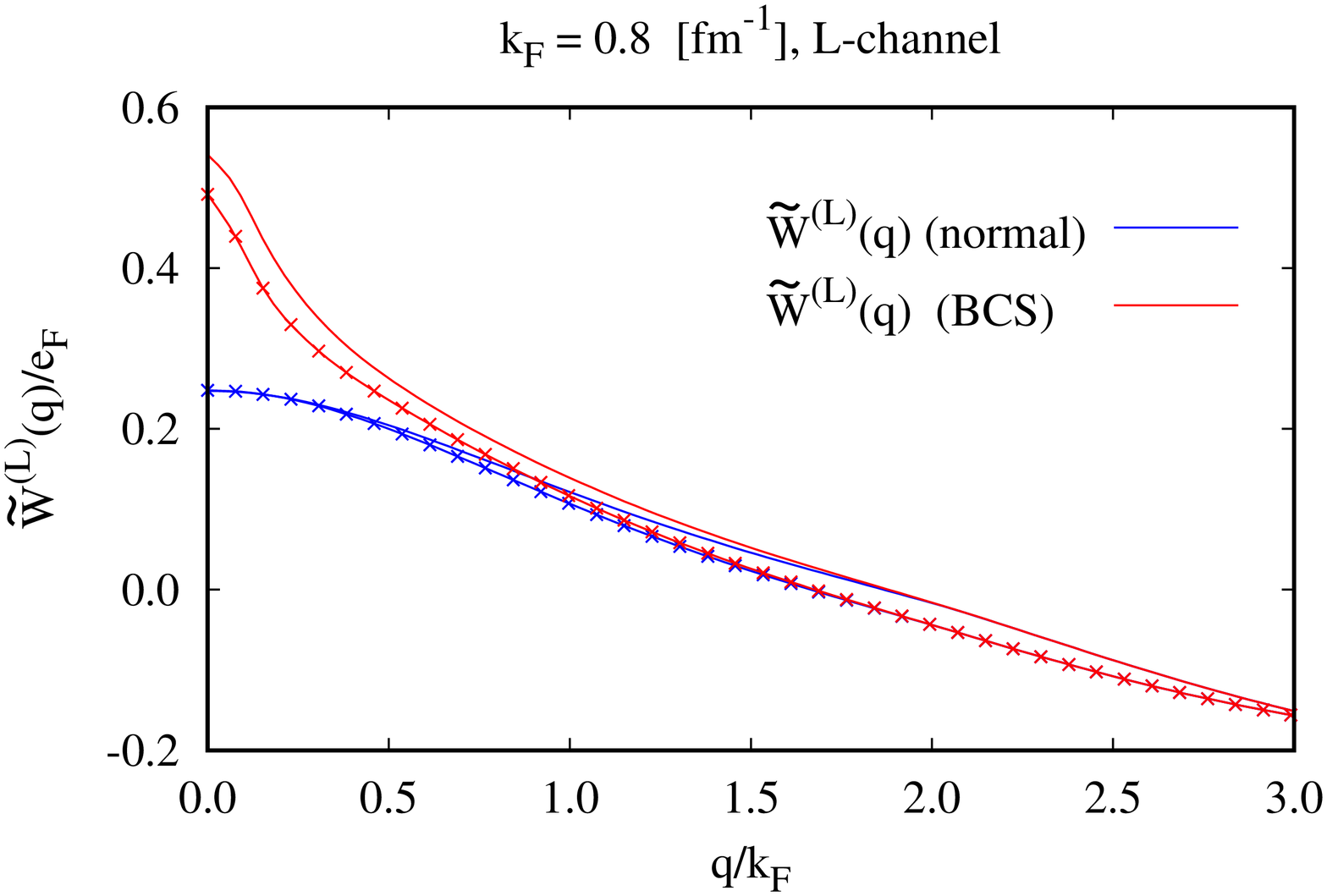}
    \caption{Same as Fig. \ref{fig:twistplot04} for
      $\KF = 0.8\,\mathrm{fm}^{-1}$.\label{fig:twistplot08}
    }
  \end{figure}
\end{widetext}

\begin{widetext}
  \begin{figure}
  \includegraphics[width=0.34\textwidth,angle=270]{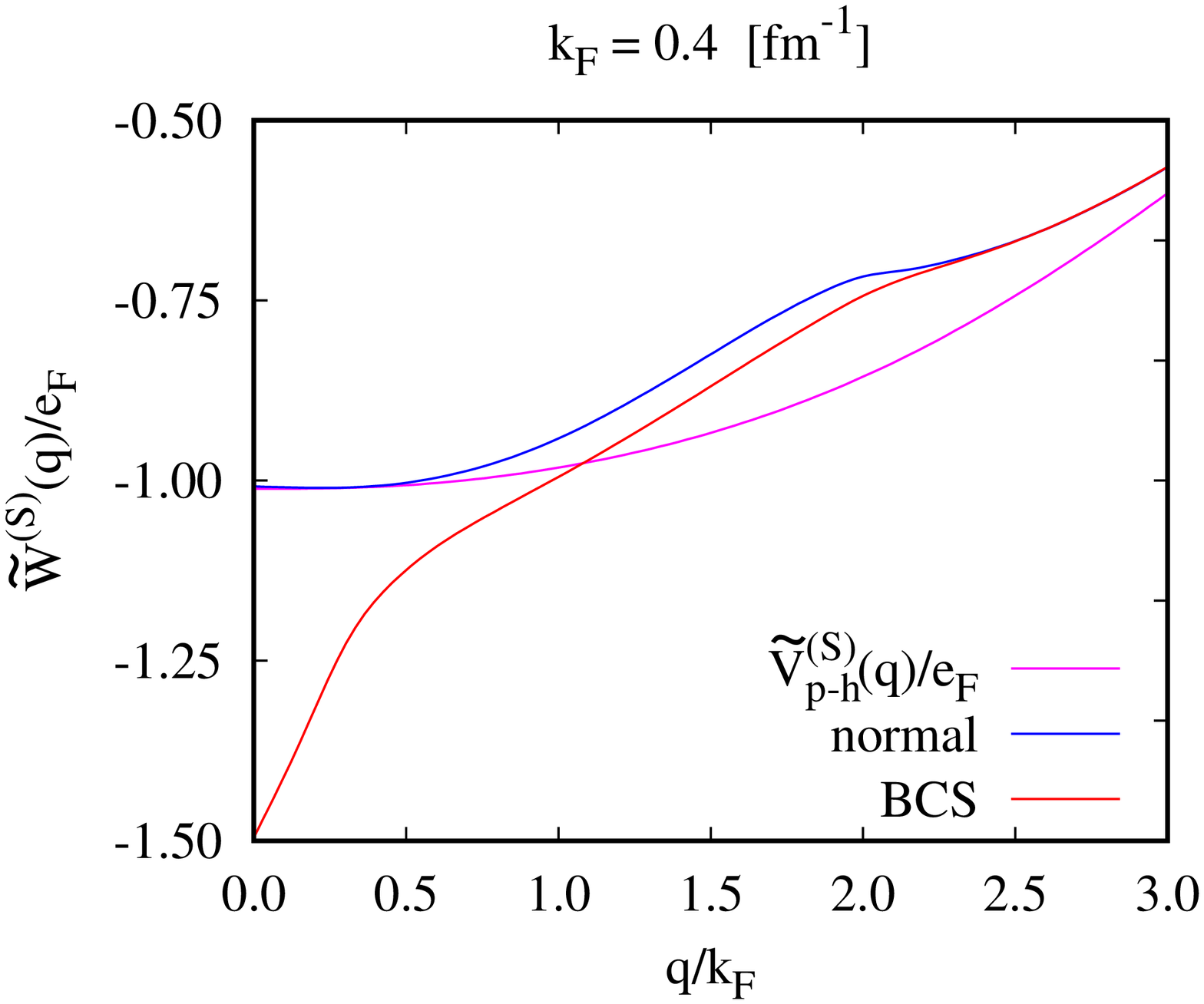}
  \includegraphics[width=0.34\textwidth,angle=270]{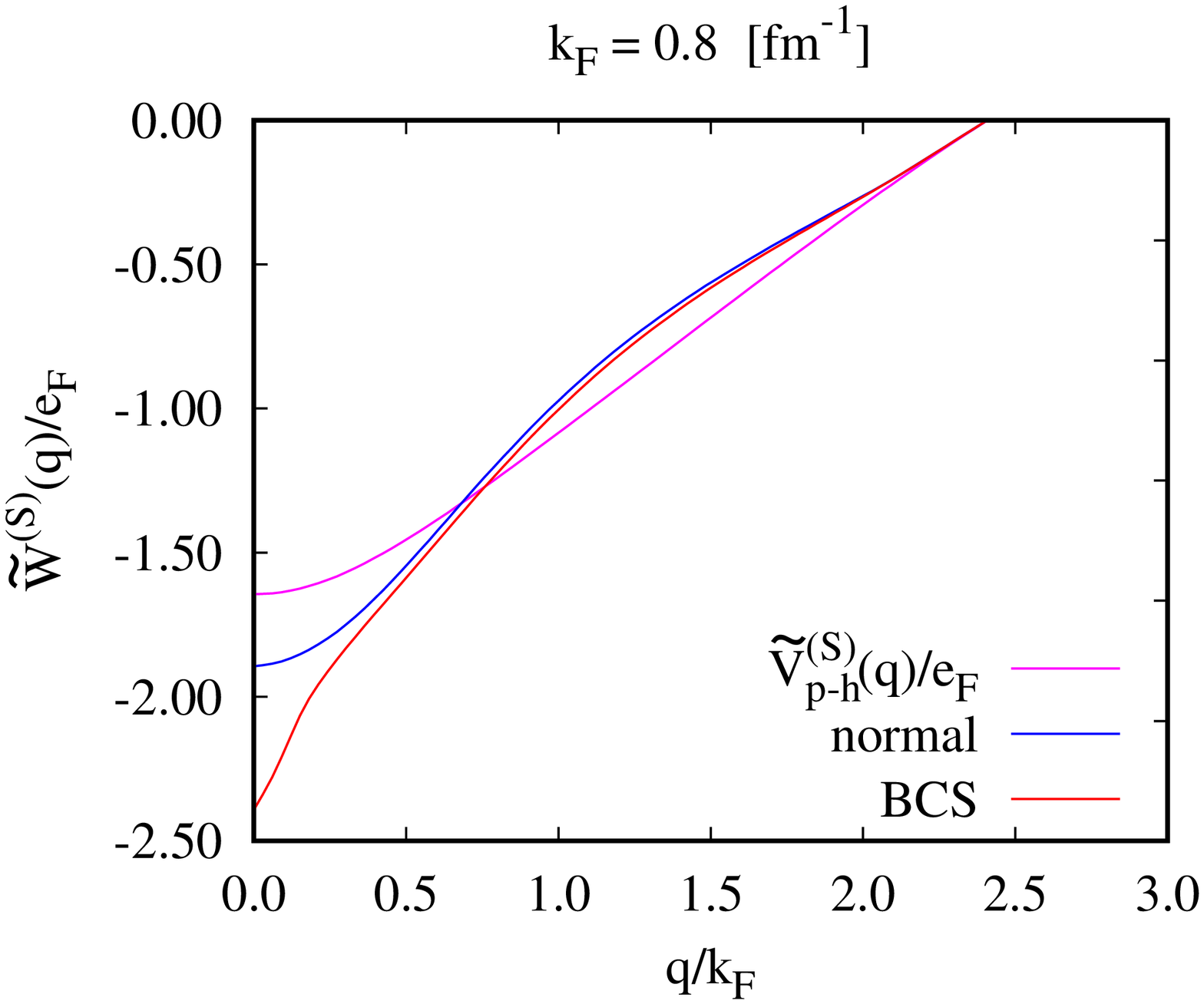}
    \caption{The left figures shows, for $\KF =
      0.4\,\mathrm{fm}^{-1}$, the singlet-$S$ wave effective
      interaction, using both the normal system Lindhard function
      (blue line) and the superfluid Lindhard function (red
      line). Also shown is the ``direct'' part of the particle-hole
      interaction (magenta line). The right figure shows the same
      potentials at $\KF = 0.8\,\mathrm{fm}^{-1}$. Both figures refer
      to the ``beyond parquet'' calculation. \label{fig:potplot} }
\end{figure}
\end{widetext}

Since we are concerned with $^1S_0$ pairing, we need to map the $\1$,
$\hat L$ and $\hat T$ channel interactions onto the $S$-wave,
\begin{equation}
  \widetildeto{\Gamma}{W}^{(S)}(q)=  \widetildeto{\Gamma}{W}^{(\1)}(q)-\widetildeto{\Gamma}{W}^{(L)}(q)-2\widetildeto{\Gamma}{W}^{(T)}(q)\,.
  \label{eq:Wsinglet}
\end{equation}
The interactions are shown in Fig. \ref{fig:potplot}. Somewhat
unexpectedly, the results show much less effect from using the
superfluid Lindhard functions.  The reason is found in the fact that
the corrections go, in both the central and the spin channels, in the
same direction and lead to an apparent partial cancellation, see
Eq. \eqref{eq:Wsinglet}. We could not see an argument that this
cancellation is generic, but rather we consider it a coincidence.

\begin{figure}
\centerline{\includegraphics[width=0.5\textwidth,angle=270]{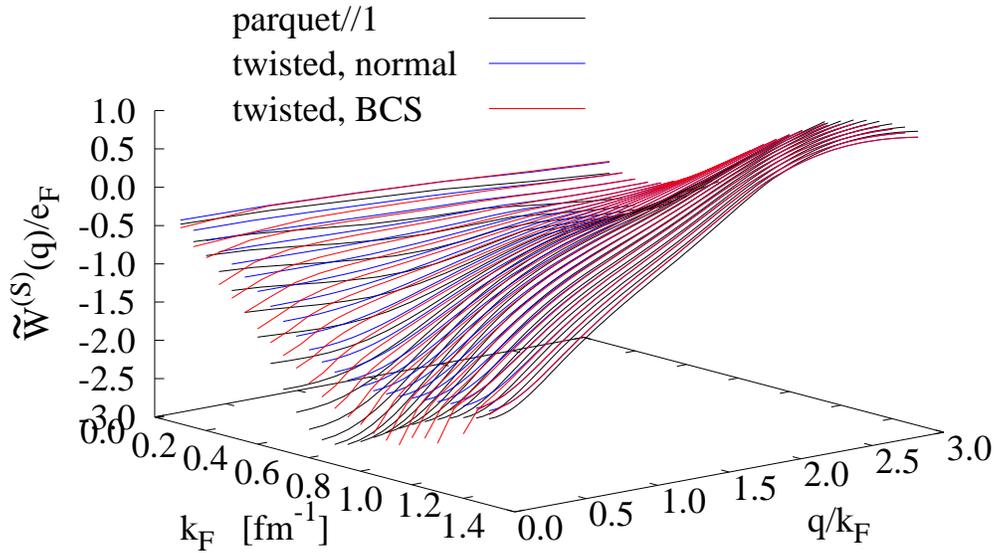}}
\caption{(color online) The figure shows the density dependence of the
  singlet pairing interaction in both ``parquet//1'' approximation
  (black lines) and including both ``beyond parquet'' corrections and
  those stemming from using a superfluid Lindhard function (red
  lines). \label{fig:WSplot_3d} }
\end{figure}

Fig. \ref{fig:WSplot_3d} gives an overall account of the density
dependence of the $S$-wave pairing interaction. Generally, the
inclusion of ``beyond-parquet'' diagrams reduces the interaction strength,
the effect is most pronounced at intermediate densities. We also see
clearly that the corrections from using a superfluid Lindhard
function are smaller, with increasing density, at longer wavelengths which is
due to the fact that the gap gets smaller.

\subsection{BCS pairing}
\label{ssec:bcs}

Once the ground-state correlations and effective interactions are
known, the superfluid gap function $\Delta_{\kvec}$ can be determined
by solving the gap equation (\ref{eq:gap}).

The gap equation was solved by the eigenvalue method with an adaptive
mesh as outlined in the appendix of Ref.~\citenum{cbcs}.  We have
adopted a free single-particle spectrum for $e_{\kvec}$ as it occurs
in Eqs.~(\ref{eq:Pdef}) and (\ref{eq:gap}).  One could also use the
actual spectrum of CBF single-particle energies \cite{CBF2}, in both
the pairing interaction (\ref{eq:Pdef}) and the denominator of
Eq.~(\ref{eq:gap}). We have discussed and studied the effect of these
modifications in previous work \cite{ectpaper}, there is no reason for
repetition. A recent very extensive comparison with earlier work
\cite{Wam93,Schulze96,Schwenk03,PhysRevC.72.054313,Cao2006,%
  PhysRevC.77.054308,Gandolfi2009b,Fabrocinipairing,Fabrocinipairing2,%
  PhysRevC.94.025802} is
found in Ref. \onlinecite{Pavlou2017}. We can, therefore, focus in
this paper on the aspect where we went beyond previous work
\cite{fullbcs,ectpaper}.

\begin{widetext}
\begin{figure}
  \includegraphics[width=0.34\columnwidth,angle=-90]{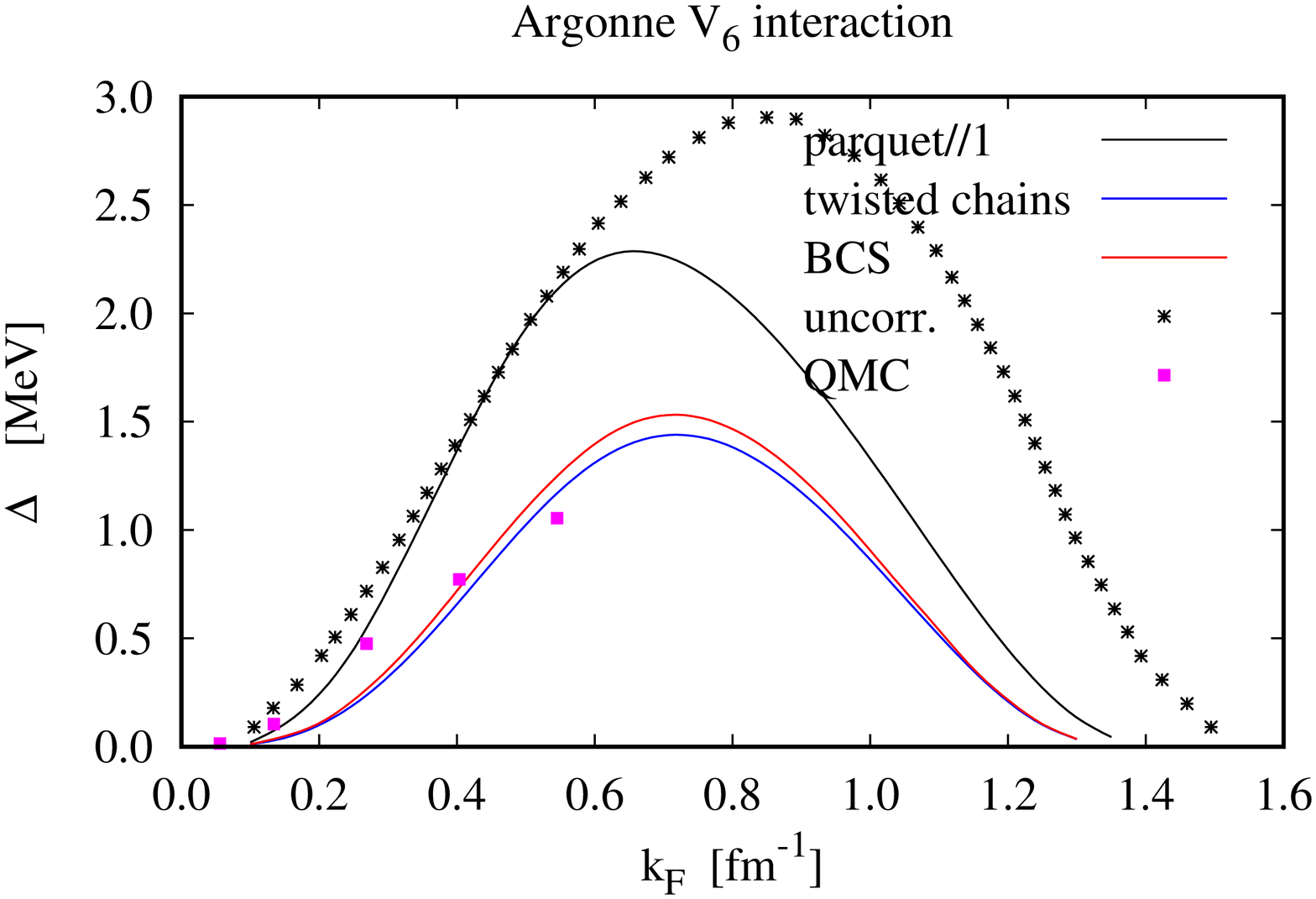}
  \includegraphics[width=0.34\columnwidth,angle=-90]{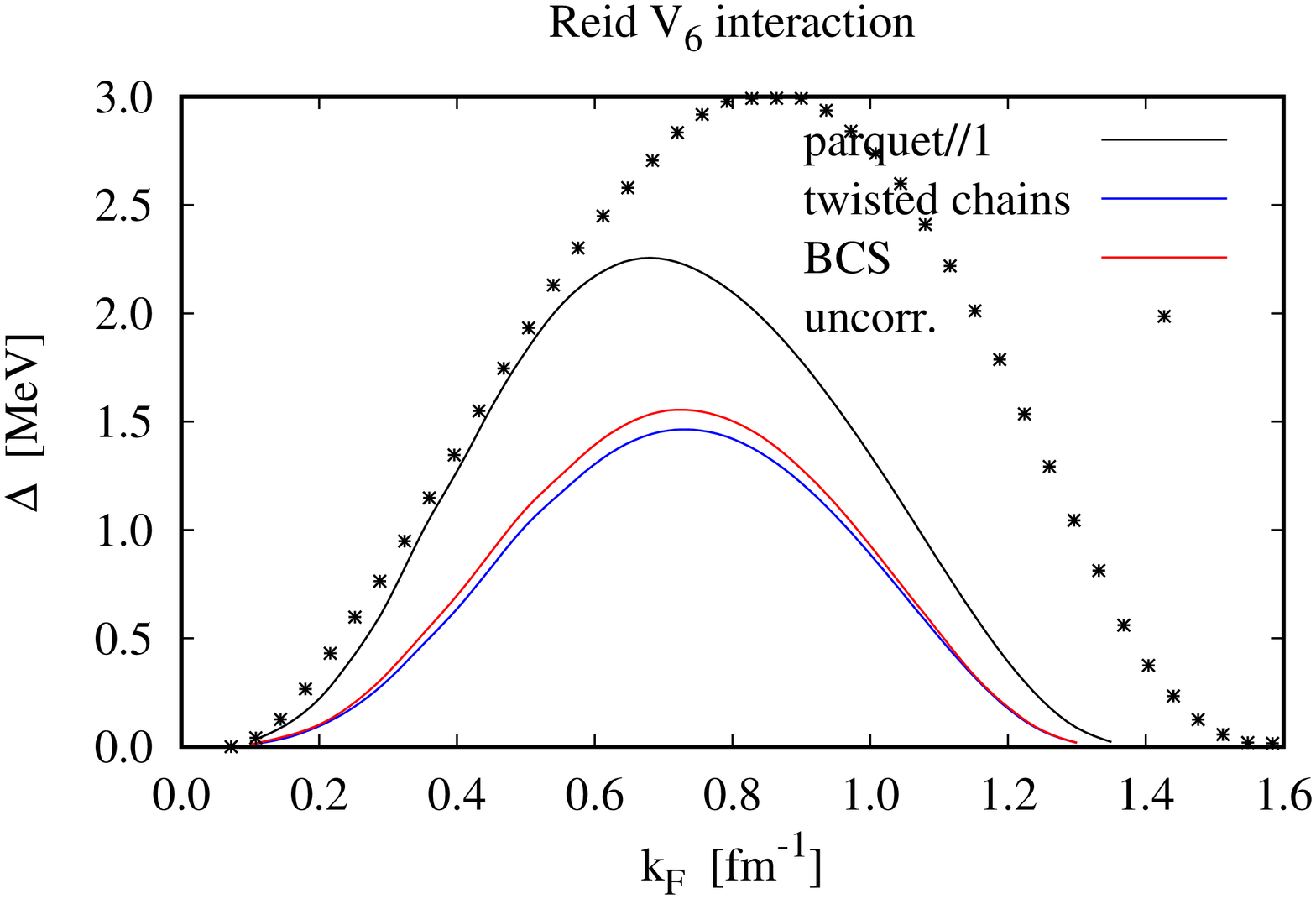}
\caption{(color online) Superfluid gap $\Delta_{\KF}$ at the Fermi
  momentum as a function of Fermi wave number $\KF$ for the Argonne
  $V_6$ interaction (left figure) and the Reid $V_6$ potential (right
  figure). We show the parquet calculation (black curve), the ``beyond
  parquet'' results (blue curve) using the Lindhard function for
  normal systems, and the ``beyond parquet'' results using the
  superfluid Lindhard function (red curves). The crosses show the
  results for the bare Argonne and Reid interactions, these data are
  from Ref. \onlinecite{KKC96}. The magenta squares in the left figure
are quantum Monte Carlo data from Ref. \onlinecite{GC2008}.}
\label{fig:gapplot}
\end{figure}
\end{widetext}

\begin{widetext}
\begin{figure}
  \includegraphics[width=0.34\columnwidth,angle=-90]{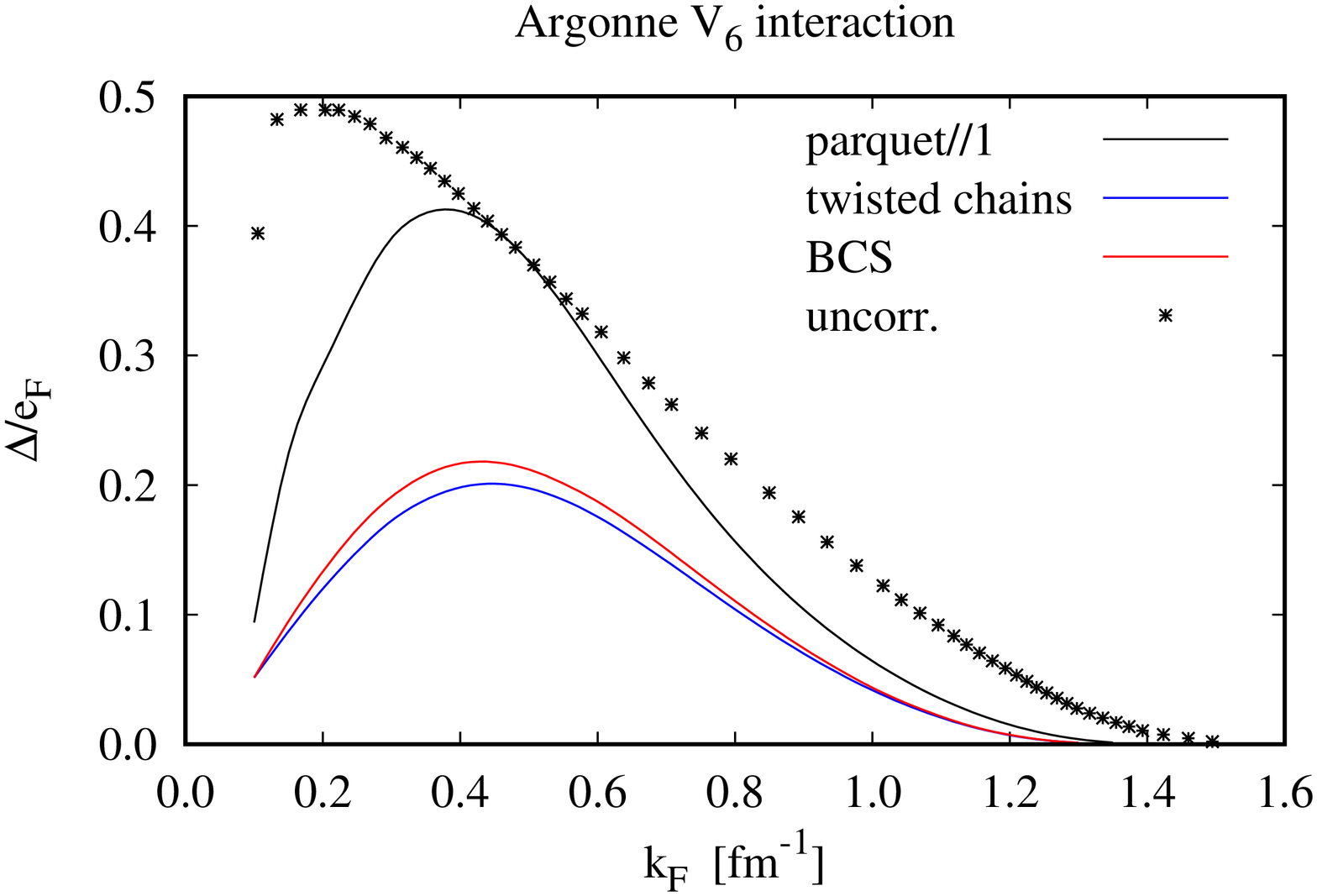}
  \includegraphics[width=0.34\columnwidth,angle=-90]{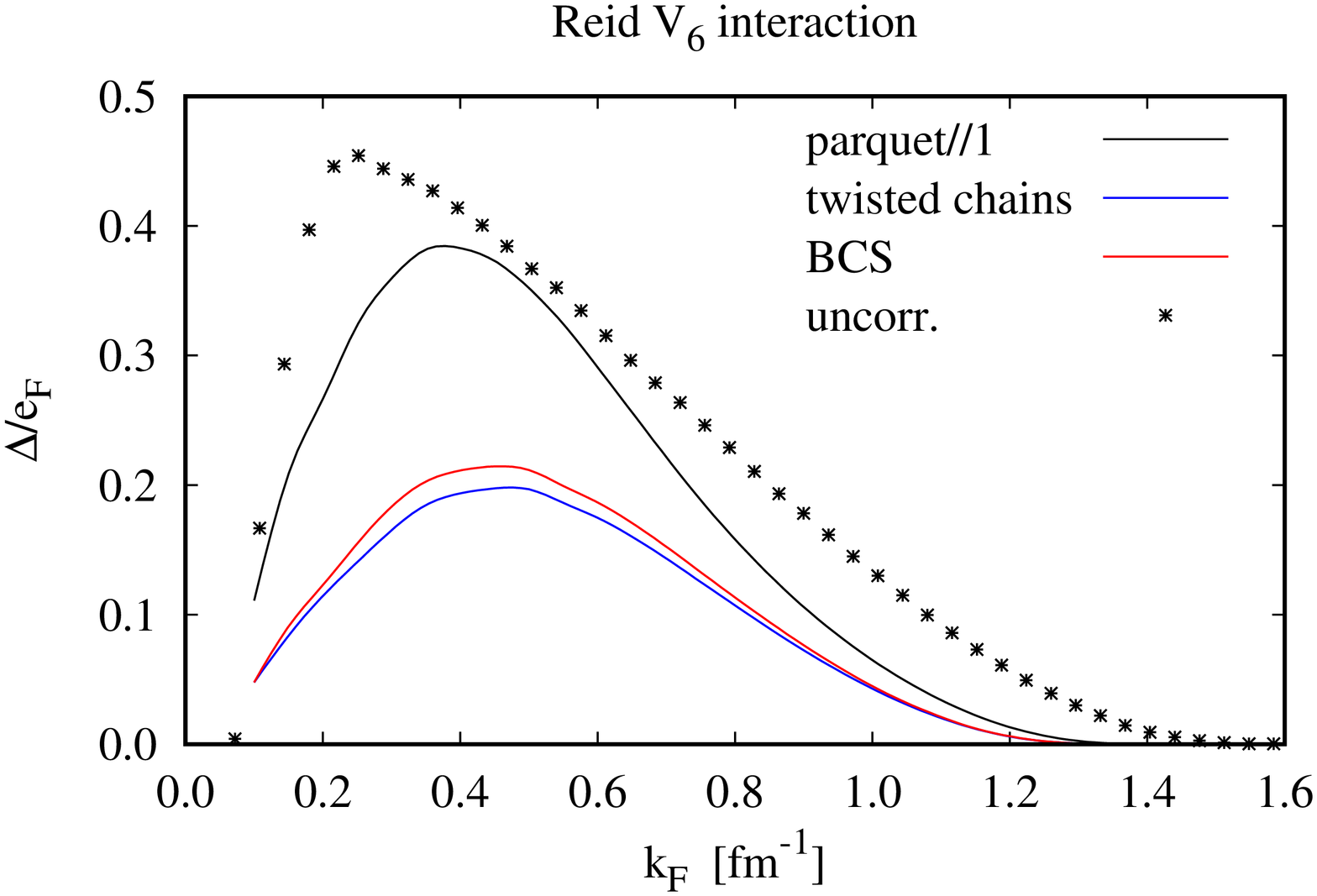}
  \caption{Same as Fig. \ref{fig:gapplot} for the gap in units of the
    Fermi energy of the non-interacting Femri gas.}
\label{fig:gafplot}
\end{figure}
\end{widetext}

Our results for the superfluid gap for the two potentials are shown in
Fig.~\ref{fig:gapplot}. Evidently the difference of the gap between
these two potential models is almost negligible and certainly within
the accuracy of both the FHNC/parquet//1 approximation. We have above
shown that specific ``beyond parquet'' corrections to the effective
interaction should enhance the repulsion between particles in the
singlet state, and Figs. \ref{fig:gapplot} and \ref{fig:gafplot} show
exactly this effect. In fact, these contributions bring our results
quite close to the quantum Monte Carlo data of
Ref. \onlinecite{GC2008}. On the other hand, the influence of using a
superfluid Lindhard function appears modest. Considering that
inspection of the individual pieces of the effective interactions
suggests exactly the opposite, we conclude that our specific results
are circumstantial and may well be totally different for other
interactions or, for example, $P$-wave pairing.

In comparison to the quantum Monte Carlo data of
Ref. \onlinecite{GC2008} it must, of course, be noted that our
interaction model is somewhat different. We have used the full $v_6$
interaction, wheras Ref. \onlinecite{GC2008} uses the $S$ wave part of
the Argonne potential. We have tried to use that interaction too, but
it turned out that the pure $S$-wave interaction leads to a spinodal
instability in which case the parquet equations have no solution.

\section{Summary and Prospects}
\label{sec:summary}

The work reported in this paper represents the most rigorous
calculation yet performed for nuclear systems within correlated BCS
theory. We have described new calculations of the pairing gap in the
$^1S_0$ partial-wave channel.

Our work goes beyond the calculations reported in
Ref. \onlinecite{ectpaper} in several important aspects: We have
replaced the state-independent FHNC/parquet summation method by the
state-dependent parquet summation method \cite{v3eos}. We have also
included the leading ``beyond parquet'' corrections \cite{v3twist},
the reason for why these diagrams can be important in particular in
the $^1S_0$ interaction channel has been discussed in the introduction
in connection with Fig. \ref{fig:vsummary}. Finally, we have used a
superfluid Lindhard function for the calculation of the particle-hole
propagator.  We found that each of these effects is quite substantial
and the fact that the sum of all of these corrections is modest seems
circumstantial rather than generic.

We need to re-iterate the importance of the energy numerator term
which originates from the fact that the correlated BCS theory is
formulated in terms of what should be considered a static
approximation of the $T$-matrix. We have examined, in
Eq. \ref{eq:BCS2ndorder}, the relationship between low-order
variational calculations and the analysis of the gap equation due to
Cooper, Mills, and Sessler (See also Ref.
\onlinecite{NozieresSchmittRink}). An important issue is the
demonstration of how Jastrow-Feenberg correlations assume the task of
short-ranged screening which is otherwise accomplished by the pair
wave function $\chi(r)$, Eq. \eqref{eq:chidef}. This works, of course,
only for state-dependent interactions if the correlations are
optimized in each operator channel separately. This casts some doubts
on earlier calculations of the superfluid gap, including our own
\cite{fullbcs,ectpaper}, which use state-independent correlation
functions.

Improvements can be sought in different ways: To look at $P$-wave
pairing, we need to extend the theory to include a spin-orbit
interaction. Another option is to add some phenomenological
information into the particle-hole interaction $\hat V_{\rm p-h}(q)$
in order to enforce the agreement between the Fermi liquid parameter
$F_0^s$ obtained from the long wavelength limit \eqref{eq:F0s} and
from the hydrodynamic derivative \eqref{eq:FermimcfromVph}. To do the
same for $F_0^a$ requires to extend the theory to arbitrary
spin-polarization. Work along these lines is in progress.

Another important aspect that we have not touched in this paper is the
importance of three-body interactions. There is the general consensus
that three-body interactions are important in nuclear systems at
higher densities. The literature on the issue is vast, see Ref.
\onlinecite{PhysRevC.89.044321} for a recent discussion. For the
problem at hand, three-body forces are expected to be most important
for P-wave pairing at high densities, see {\em e.g.} Ref.
\onlinecite{PhysRevC.78.015805} and Ref. \onlinecite{JLTP189_361} for
a very complete discussion of the earlier literature and, in
particular, the sensitivity of the pairing gap on the choice of the
interaction. A generalization of parquet theory for three-body forces
could be carried out along the lines of three-body Jastrow-Feenberg
\cite{Woo3body1,Woo3body2} or parquet theory \cite{TripletParquet} but
has not been carried out so far.  We hesitate very much to speculate
what the effect of including three-body forces in parquet/CBF theory
would do, a solid calculation would go far beyond the scope of this
paper.

\appendix
\section{Correlated Basis Functions Theory}
\label{app:CBF}

For the development of a microscopic theory for superfluid systems we
need the basic ingredients of correlated basis functions (CBF)
theory. We give here only the definitions of the relevant quantities
to the extent that they are needed for the present work, details may
be found in pedagogical material \cite{KroTrieste} and review articles
\cite{Johnreview,polish}.

CBF theory uses the correlation operator $F$ to generate a complete
set of correlated and normalized $N$-particle basis states through
\begin{equation}
\ket{\Psi_{\bf m}^{(N)}} =
\frac{F_{\!N} \; \ket {{\bf m}^{(N)} } }
{\bra{{\bf m}^{(N)}}  F_{\!N}^{\dagger} F^{\phantom{\dagger}}_{\!N}
\ket{{\bf m}^{(N)}}
\rangle^{1/2} } \;,
\label{eq:States}
\end{equation}
where the $\{\ket {{\bf m}^{(N)}}\}$ form a complete basis of
model states, normally consisting of Slater determinants of single
particle orbitals.

In general, we label ``hole'' states which are occupied in $
\ket{{\bf o}}$ by $h$, $h'$, $h_i\;, \ldots\,$, and unoccupied
``particle'' states by $p$, $p'$, $p_i\;,$ \textit{etc.}.  To display
the particle-hole pairs explicitly, we will alternatively to the
notation $\ket{{\bf m}}$ use
$\ket{\Psi_{p_1 \ldots p_d\, h_1\ldots h_d}} $.  A basis
state with $d$ particle-hole pairs is then
\begin{equation}
\ket{\Psi_{p_1 \ldots p_d\, h_1\ldots h_d}} 
=\left[I_{p_1,\dots h_1}^{(N)}\right]^{-1/2}F_N
\creat{p_1}\cdots\creat{p_d}\annil{h_d}\cdots\annil{h_1}\ket{\bf o}
\,.
\label{eq:psimph}
\end{equation}

For the off-diagonal elements $O_{\bf m,n}$ of an operator $\hat O$, we
sort the quantum numbers $m_i$ and $n_i$ such that $\ket{{\bf m}}$
  is mapped onto $\ket{{\bf n}}$ by
\begin{equation}
\label{eq:defwave}
\ket{{\bf m}} = \creat{m_1}\creat{m_2}
\cdots 
\creat{m_d} \; \annil{n_d} \cdots \annil{n_2}\annil{n_1}  
\ket{{\bf n}} \;.
\end{equation}
From this we recognize that, to leading order in the particle number
$N$, any matrix element of an operator $\hat O$
\begin{equation}
  O_{\bf m,n} = \bra{\Psi_{\bf m}} \hat O \ket{\Psi_{\bf n}}
\end{equation}
depends only on the {\em difference\/} between the states
$\ket{{\bf m}}$ and $\ket{{\bf n}}$, and {\em
not\/} on the states as a whole.  Consequently, $O_{\bf m,n}$ can be
written as matrix element of a $d$-body operator
\begin{equation}
\label{eq:defmatrix}
O_{\bf m,n} \equiv \bra{ m_1\, m_2 \, \ldots m_d \,} 
{\cal O}(1,2,\ldots d) \,\ket{n_1\,
n_2 \, \ldots n_d}_a \;.
\end{equation}
(The index $a$ indicates antisymmetrization.) 

The key quantities for the execution of the theory are diagonal and
off-diagonal matrix elements of unity and $H\!-\!H_{\bf o}$,
\begin{eqnarray}
M_{\bf m,n} &=& \ovlp{\Psi_{\bf m}}{\Psi_{\bf n}}
\equiv \delta_{\bf m,n} +  N_{\bf m,n}\;,
\label{eq:defineNM}
\\
W_{\bf m,n} &=& \bra{\Psi_{\bf m}}H-\frac{1}{2}\left(H_{\bf m}+H_{\bf n}\right)\ket{\Psi_{\bf m}} \,.
\label{eq:defineW}
\end{eqnarray}
Eq. (\ref{eq:defineW}) defines a natural decomposition
\cite{CBF2,KroTrieste} of the matrix elements of $H$ into the
off-diagonal quantities $W_{\bf m,n}$ and $N_{\bf m,n}$ and diagonal
quantities $H_{\bf m}$.

To leading order in the particle number, the {\em diagonal\/} matrix
elements of $ H\!-\!H_{\bf o}$ become additive, so
that for the above $d$-pair state we can define the CBF single
particle energies
\begin{equation}
\bra{\Psi_{\bf m}} H\!-\!H_{\bf o}  \ket{\Psi_{\bf m}} \>\equiv\>
\sum_{i=1}^d e_{p_ih_i}  + {\cal O}(N^{-1}) \;,
\label{eq:CBFph}
\end{equation}
with $e_{ph} = e_p - e_h$ where
\begin{align}
e_p &=\phantom{-}\bra{\Psi_p}\,\ H\!-\!H_{\bf o}
\ket{\Psi_p} = t(p) + u(p)\nonumber\\
e_h &=-\bra{\Psi_h}\,H\!-\!H_{\bf o}\ket{\Psi_h}\ = t(h) + u(h)\,
\label{eq:spectrum}
\end{align}
and $u(p)$ is an average field that can be expressed in terms of the
compound diagrammatic quantities of FHNC theory \cite{CBF2}.

According to (\ref{eq:defmatrix}),
$W_{{\bf m},{\bf n}}$  and $N_{{\bf m},{\bf n}}$ define 
$d-$particle operators ${\cal N}$ and ${\cal W}$, {\em e.g.\/}
\begin{eqnarray}
N_{{\bf m},{\bf o}} &\equiv& N_{p_1p_2\ldots p_d \,h_1h_2\ldots h_d,0} \nonumber\\
&\equiv& \bra{ p_1p_2\ldots p_d \,}\, {\cal N}(1,2,\ldots,d)\,
\ket{\,h_1h_2\ldots h_d }_a  \;,\nonumber\\
W_{{\bf m},{\bf o}} &\equiv& W_{p_1p_2\ldots p_d \,h_1h_2\ldots h_d,0}\nonumber\\
&\equiv&  \bra{ p_1p_2\ldots p_d \,}\, {\cal W}(1,2,\ldots,d)\,
\ket{\,h_1h_2\ldots h_d }_a  \;.\qquad\;
\label{eq:NWop}
\end{eqnarray}
Diagrammatic representations of ${\cal N}(1,2,\ldots,d)$ and ${\cal
W}(1,2,\ldots,d)$ have the same topology \cite{CBF2}.  In
homogeneous systems, the continuous parts of the $p_i$, $h_i$ are wave
numbers ${\bf p}_i$, ${\bf h}_i$; we abbreviate their difference as ${\bf
q}_i$.

In principle, the ${\cal N}(1,2,\ldots,d)$ and ${\cal
  W}(1,2,\ldots,d)$ are non-local $d$-body operators.  Above, we have
shown that we need, for examining pairing phenomena, only the two-body
operators. Moreover, the low density of the systems we are examining
permits the same simplifications of the FHNC theory that we have
spelled out in Sec. \ref{ssec:FHNC}. In that approximation, the
operators ${\cal N}(1,2)$ and ${\cal W}(1,2)$ are local, and we have
\cite{polish}
\begin{eqnarray}
{\cal N}(1,2) &=& {\cal N}(r_{12}) = \Gamma_{\!\rm dd}(r_{12})\nonumber\\
{\cal W}(1,2) &=& {\cal W}(r_{12})\,,\quad \tilde {\cal W}(k) \equiv
\widetildeto{\Gamma}{W}(k) = - \frac{t(k)}{\SF(k)}\tilde \Gamma_{\!\rm dd}(k)\,.
\label{eq:NWlocApp}
\end{eqnarray}

\acknowledgments Encouragement for this work was derived from a
workshop on {\em Nuclear Many-Body Theories: Beyond the mean field
  approaches\/} at the Asia Pacific Center for Theoretical Physics in
Pohang, South Korea, in July 2019 and the winter school on {\em
  Superfluidity and Transport for Multimessenger Physics of Compact
  Stars} in Karpacz, Poland, Feb. 2020.

\pagebreak

\bibliography{papers}
\bibliographystyle{apsrev4-1}

\end{document}